\documentclass[preprint,floats,tightenlines,11pt,eqsecnum,showpacs,aps]{revtex4}
\usepackage{amsmath}
\usepackage{graphicx}
\begin{document}
\def\appls{\hbox{$<$\kern-.75em\lower 1.00ex\hbox{$\sim$}}}


\title{CONSISTENCY OF $\pi\pi$ PHASE SHIFT ANALYSES WITH\\$\rho^0(770)-f_0(980)$ SPIN MIXING IN $\pi^- p \to \pi^-\pi^+n$}

\author{Miloslav Svec\footnote{electronic address: svec@hep.physics.mcgill.ca}}
\affiliation{Physics Department, Dawson College, Montreal, Quebec, Canada H3Z 1A4}
\date{June 8, 2015}

\begin{abstract}

We have performed two analytical $\pi\pi$ phase-shift analyses using a Standard amplitude analysis of the CERN data on $\pi^- p \to \pi^-\pi^-n$ on polarized target at 17.2 GeV/c and a more recent analysis of the same data using a Spin Mixing Mechanism (SMM). There are two solutions for helicity amplitudes labeled (1,1) and (2,2) in the Standard analysis and SpinMixing and S-Matrix in the SMM analysis which are related to $\pi\pi$ scattering amplitudes. Our first phase shift analysis is an elastic scattering analysis below $K\bar{K}$ threshold. Our second analysis is a joint $\pi\pi$ phase shift analysis of $\pi^-\pi^+$ and $\pi^0\pi^0$ data below 1080 MeV. Our elastic Solution (2,2)1 and the elastic Solution Spin Mixing 1 for $\delta^0_S$ are in a remarkable agreement with the 1997 Cracow Solution Down-flat using the same CERN data on polarized target. Our joint Solution (2,2) and joint Solution Spin Mixing are also in a remarkable agreement with the 2002 joint Cracow Solution Down-flat. Solutions elastic (1,1) and joint (1,1) agree with the Cracow Solutions Up-flat and are rejected. 

Model independent amplitude analyses of measurements on polarized targets of $\pi^- p \to \pi^-\pi^-n$ at 17.2 and 1.78 GeV/c and $\pi^+ n \to \pi^+\pi^-p$ at 5.98 and 11.85 GeV/c reveal evidence for $\rho^0(770)-f_0(980)$ spin mixing in the $S$-wave transversity amplitudes. The transversity amplitudes define single-flip helicity amplitudes which have been related to $\pi\pi$ scattering amplitudes. In all our and Cracow solutions the phase-shift $\delta^0_S$ passes through $90^\circ$ at or near 770 MeV hinting at $\rho^0(770)-f_0(980)$ mixing. There is no evidence for such mixing in the Solutions S-Matrix as expected from the S-matrix $\pi\pi$ scattering amplitudes. All four Solutions $S$-Matrix are very similar suggesting the existence of a unique solution for the phase-shift $\delta^0_S$ and lending credence to their interpretation as genuine $S$-matrix amplitudes. Our key observation is that all these solutions for $\delta^0_S$ are consistent with the evidence for $\rho^0(770)-f_0(980)$ spin mixing in the measured transversity amplitudes from which all these phase shifts ultimately arise.

\end{abstract}

\maketitle

\tableofcontents

\newpage
\section{Introduction.}

In Quantum Field Theory particle scattering and decay processes are isolated and time reversible events governed by $S$-matrix unitary evolution law
\begin{equation}
\rho_f=S \rho_i S^+
\end{equation}
which evolves pure initial states $\rho_i$ into pure final states $\rho_f$. In $\pi N \to \pi \pi N$ processes the unitary evolution of pure states to pure states predicts that the relative phase between any two unnatural exchange and between any two natural exchange partial wave transversity amplitudes must be $0^\circ$ or $\pm 180^\circ$~\cite{svec13a,svec13b}. This prediction is in a complete disagreement with experimentally determined relative phases in amplitude analyses of all measurements on polarized targets: $\pi^- p\to \pi^-\pi^+ n$ at 17.2 GeV/c~\cite{becker79a,becker79b,chabaud83,rybicki85,kaminski97} and at 1.78 GeV/c~\cite{alekseev98,alekseev99}, and $\pi^+ n \to \pi^+\pi^- n$ at 5.98 and 11.85 GeV/c~\cite{lesquen85,svec92a,svec96,svec97a}. 

The sharp contrast between the predicted unitary phases and the observed phases is an unambigous evidence for a nonunitary evolution of the final state $\rho_f(S)$ produced by the $S$-matrix dynamics into the observed final state $\rho_f(O)$ arising from the pure dephasing interaction of the state $\rho_f(S)$ with a quantum environment~\cite{svec13a}. In pure dephasing interactions all four-momenta and the identities of all final state particles do not change and there is no exchange of four-momentum between the produced final state and the environment. There is no interaction with the quantum environment in two-body scattering or decays.

The consistency of this nonunitary interaction with the Standard Model and its conservation laws predicts that in $\pi^- p\to \pi^-\pi^+ n$ the observed partial wave transversity amplitudes are a unitary transform of $S$-matrix transversity amplitudes with dipion spins $J_1$ and $J_2$ with $|J_2-J_1|=1$ and the same dipion helicity $\lambda$~\cite{svec13b}. We refer to this transform as a spin mixing mechanism (SMM). There is no spin mixing in $\pi^- p\to \pi^0\pi^0 n$ and $\pi^+ p\to \pi^+\pi^+ n$. For the $S$- and $P$-wave amplitudes the spin mixing mechanism reads~\cite{svec13b}
\begin{eqnarray}
L_\tau & = & + e^{i\phi} \bigl ( +\cos \theta S_\tau^0 + e^{i\phi} \sin\theta 
L_\tau^0 \bigr )\\
\nonumber
S_\tau & = & +e^{i\phi} \bigl ( -\sin \theta S_\tau^0 + e^{i\phi} \cos\theta 
L_\tau^0 \bigr )
\end{eqnarray}
where $S_\tau, L_\tau, \tau=u,d$ are the observed spin mixing amplitudes, $S_\tau^0, L_\tau^0, \tau=u,d$ are the $S$-matrix amplitudes and where $\theta$ and $\phi$ are spin mixing parameters. Here $\tau=u,d$ stand for the target nucleon transversity with spin $"up"$ and $"down"$, respectively. Thus the consistency of the pure dephasing interaction with Standard Model alone predicts $\rho^0(770)-f_0(980)$ spin mixing in the observed amplitudes. Such spin mixing is forbidden in the $S$-matrix amplitudes by the Lorentz symmetry of the $S$-matrix. While the final state $\rho_f(S)$ is produced by the Lorentz symetric dynamics of the Standard Model, the non-unitary evolution of this state into $\rho_f(O)$ gives rise to the spontaneous violation of rotational/Lorentz symmetry in the observed amplitudes. 

Evidence for a rho-like state (later called $\sigma(750))$ in the $S$-wave in $\pi^- p\to \pi^-\pi^+ n$ at low energies dates back to  1960's~\cite{hagopian63,islam64,patil64,durand65,baton65}. A review of these first analyses is given in~\cite{gasiorowicz66}. A rho-like state in the $S$-wave was suggested by the early analyses~\cite{donohue79,svec92c} of the CERN measurements of $\pi^- p\to \pi^-\pi^+ n$ on the polarized target at 17.2 GeV/c. Amplitude analyses of these measurements~\cite{becker79a,becker79b,chabaud83,rybicki85,kaminski97,svec96,svec97a}, ITEP measurements at 1.78 GeV/c~\cite{alekseev98,alekseev99} and CERN-Saclay measurements of $\pi^+ n\to \pi^+\pi^- p$ at 5.98 and 11.85 GeV/c~\cite{svec96,svec97a} confirmed the existence of this rho-like state. A comprehesive survey of all this evidence is presented in Ref.~\cite{svec12d}.

These findings were controversial because the measurements of $\pi^- p \to \pi^0 \pi^0 n$ at CERN in 1972~\cite{apel72} and at BNL in 2001~\cite{gunter01,pi0pi0pwa} found no evidence for the rho-like meson in the $S$-wave amplitudes. Using three different methods we show in a recent work~\cite{svec12b} that the rho-like resonance in the $S$-wave transversity amplitudes arises entirely from the contribution of the $\rho^0(770)$ resonance. In addition, there is a dip at the $f_0(980)$ mass in the $P$-wave aplitude $|L_d|^2$. These results present evidence for a $\rho^0(770)-f_0(980)$ mixing in $\pi^- p \to \pi^- \pi^+n$. Since there is no $P$-wave in $\pi^- p \to \pi^0 \pi^0 n$ this explains why there is no rho-like resonance observed in this process. 

The question arises what is the quantum environment. In Ref.~\cite{svec13a} we propose to identify it with a component of Dark Matter and elaborate on this conjecture in Ref.~\cite{svec14a}. The pure dephasing interactions of the produced final state $\rho_f(S)$ with Dark Matter are not rare events but require high statistics measurements on polarized target for their detection. In this picture the signature of Dark Matter is spin mixing or the violation of certain phase conditions by the observed amplitudes~\cite{svec13b}.

In Analysis I of Ref.~\cite{svec12b} we determine four solutions labeled $(i,j)$, $i,j=1,2$ for the moduli $|S_u(i)|^2,|S_d(j)|^2$, $i,j=1,2$. The Solutions $(1,1)$ and $(2,2)$ correspond to the Solutions "Up" and "Down", respectively, of the 1997 analysis of Kami\'{n}ski, Le\'{s}niak and Rybicki~\cite{kaminski97} labeled $\chi^2$ $'97$. These two analyses share the same data set~\cite{rybicki96}. Both Solutions $|S_d(j)|^2, j=1,2$ resonate at $\rho^0(770)$ mass in both our Analysis I and KLR $97$. We shall refer to these amplitude analyses as "Standard" analyses.

In our latest amplitude analysis~\cite{svec14a} of the CERN data set~\cite{rybicki96} we use spin mixing mechanism to determine anew the spin mixing amplitudes $(1,1)$ and $(2,2)$, the corresponding $S$-matrix amplitudes and the spin mixing parameters $\theta$ and $\phi$. Spin mixing mechanism excludes the Solution $(1,1)$. To distinguish the accepted Solution $(2,2)$ we label it Solution SpinMixing. The corresponding solution for the $S$-matrix amplitudes we label Solution S-Matrix. We refer to this amplitude analysis as "SMM Analysis".

$\pi\pi$ scattering partial wave amplitudes $f_J(m)$ are related to the residues of the pion pole exchange in the single flip helicity production amplitudes $F^J_1(s,t,m)$ with dipion helicity $\lambda=0$~\cite{martin76,petersen77}. The $\pi \pi$ scattering amplitudes can be determined using pion exchange dominance approximation of helicity amplitudes. The required single flip helicity amplitudes can be determined from the measured transversity amplitudes provided a relative phase 
\begin{equation}
\omega=\Phi(S_d)-\Phi(S_u)
\end{equation}
between $S$-wave amplitudes of opposite transversities is known. This phase is not measured in experiments on polarized targets but it can be determined analytically from a self-consistency condition of the bilinear terms of $S$ and $P$ wave transversity amplitudes~\cite{svec12b}. At low momentum transfers $t$ the phase $\omega=\pm180^\circ$~\cite{svec12b}.

High statistics CERN-Munich data on $\pi^- p \to \pi^- \pi^+ n$ at 17.2 GeV/c on unpolarized target~\cite{grayer74} were analysed to determine $\pi\pi$ phase- 
shifts~\cite{hyams73,estabrooks73,estabrooks74,bugg96} using several model dependent methods to extract the single flip helicity amplitudes from the data. 
First $\pi\pi$ phase-shift analysis using CERN-Cracow-Munich (CCM) data on $\pi^- p \to \pi^- \pi^+ n$ at 17.2 GeV/c on polarized target was reported 1997 in Ref.~\cite{kaminski97} (henceforth referred to as KLR 97). It was revisited 2002~\cite{kaminski02} (henceforth referred to as KLR 02) in a joint analysis of the CCM $\pi^-\pi^+$ data and the E852 $\pi^0\pi^0$ data~\cite{gunter01,pi0pi0pwa}. These two analyses used an Ansatz for $\omega$ in terms of relative phases between the $S$-and $P$-wave transversity amplitudes $S_\tau$ and $L_\tau$. In contrast, in Ref.~\cite{svec12b} we determined exact $S$-wave and $P$-wave non-flip and single-flip helicity amplitudes for the Solutions $(1,1)$ and $(2,2)$, and in Ref.~\cite{svec14a} for the Solution SpinMixing and Solution S-Matrix.

In Section II we summarize the $S$- and $P$-wave transversity and helicity amplitudes in the Standard and SMM analyses. Analytical solutions for the phase-shift $\delta^0_S$ in elastic $\pi^-\pi^+$ scattering below $K\bar{K}$ threshold are presented in Section III. There are two solutions 1 and 2 for each input helicity amplitudes. The physical Solutions (2,2)1 and SpinMixing 1 pass through $90^\circ$ at 770 MeV hinting at $\rho^0(770)-f_0(980)$ mixing and are in excellent agreement with the 1997 Solution Down-flat in KLR 97. The two Solutions S-matrix 1 and S-matrix 2 are nearly flat with no hint of $\rho^0(770)-f_0(980)$ mixing and are nearly equal.

In Section IV we present our joint analysis of the CCM $\pi^-\pi^+$ data and the E852 $\pi^0\pi^0$ data below 1080 MeV. There is a unique analytical solution for $\delta^0_S$ and inelasticity $\eta^0_S$ for each input helicity amplitudes which depends on a vertex correction parameter $C_S^2$. We determine its approximate value at each mass $m$ from a certain independent condition. The Solutions $(1,1)$ joint and $(2,2)$ joint for $\delta^0_S$ are in excellent agreement with the 2002 Solutions Up-flat and Down-flat in KLR $02$, respectively. Our inelasticities are below 1 and show less variance than those in KLR $02$. The Solution SpinMixing joint is also in excellent agreement with the Solution Down-flat but shows larger errors. The Solution S-Matrix joint is again small and nearly flat below $K\bar{K}$ threshold but rises above it. Unfortunately it suffers from somewhat larger errors.

In Section V we compare in detail the method of analysis used by the Cracow group and our method. We present our conclusions in the Section VI. The elastic and joint analyses are both consistent with a weak dependence of the phase shift $\delta^0_S$ on the phase $\omega$ while the inelasticities $\eta^0_S$ show a stronger dependence. Our central conclusion is that all $\pi\pi$ phase shift analyses - our elastic and joint as well as KLR $97$ and KLR $02$ - are consistent with the observation of the $\rho^0(770)-f_0(980)$ spin mixing in the transversity amplitudes from which these phase shifts ultimately arise.

\section{Amplitude analyses of CERN measurements of $\pi^- p \to \pi^- \pi^+ n$ \\on polarized target.}

\subsection{Transversity amplitudes}

The 1997 amplitude analysis of Kami\'{n}ski, Le\'{s}niak and Rybicki~\cite{kaminski97} labeled $\chi^2$ $'97$ is based on a $\chi^2$ fit of the transversity amplitudes to the measured data at each mass bin. Our amplitude Analysis I of Ref.~\cite{svec12b} is based on a Monte Carlo solution of the analytical equations for the transversity amplitudes in each mass bin. The Analysis I used 1 million data sampling of the data error volume. Analysis using 10 million data sampling yields identical amplitudes. Both amplitude analyses share the same data set~\cite{rybicki96}. 

There are two solutions for the $S$- and $P$-wave amplitudes $S_\tau(i), L_\tau(i)$, $i=1,2$ for each transversity corresponding to Solutions Up (i=1) and Down (i=2) in the notation of Ref.~\cite{kaminski97}. For each solution the authors present the $S$-wave intensity $I(S)=|S_u|^2+|S_d|^2$ and the ratio $R=|S_u|/|S_d|$. From this data it is a simple matter to calculate the moduli of the $S$-wave transversity amplitudes. The results are shown in the Figure 1. Figure 2 shows the $S$-wave transversity amplitudes of our Analysis I. The results of the two analysis are nearly identical. In particular, they both show a clear $\rho^0(770)$ signal at 770 MeV in both Solutions 1/Up and 2/Down for the amplitude $|S_d|^2$. This is not suprising since both analyses use the same data set~\cite{rybicki96} and their methods of analyses are both legitimate methods.

In our latest amplitude analysis~\cite{svec14a} of the data set~\cite{rybicki96} we go beyond the Standard amplitude analysis and employ the Spin Mixing Mechanism (SMM) (1.2) to determine the spin mixing and the corresponding $S$-matrix transversity amplitudes. This SMM analysis also yields two Solutions for the spin mixing and the $S$-matrix amplitudes which are shown in Figure 3 and Figure 4, respectively. We see in Figure 3 that both Solutions for $|S_d|^2$ resonate clearly at 770 MeV. Moreover, there is no structure at 930 MeV in Solution 2 of $|S_d|^2$ seen in the Standard Analysis I and $\chi^2$ $'97$. However Figure 4 reveals a large difference between Solution 1 and Solution 2 for the $S$-matrix amplitude $|S^0_d|^2$. Solution 1 clearly resonates at 770 MeV with a very pronounced $\rho^0(770)$ signal. Since there can be no spin mixing in the $S$-matrix amplitudes this Solution must be rejected. In contrast, the amplitude $|S^0_d|^2$ in Solution 2 is small and nearly flat below $K\bar{K}$ threshold as expected from the non-resonating $S$-matrix amplitude. We conclude that there is a unique Solution 2 for the spin mixing and $S$-matrix transversity amplitudes in the SMM amplitude analysis.    
\begin{figure} [htp]
\includegraphics[width=12cm,height=10.5cm]{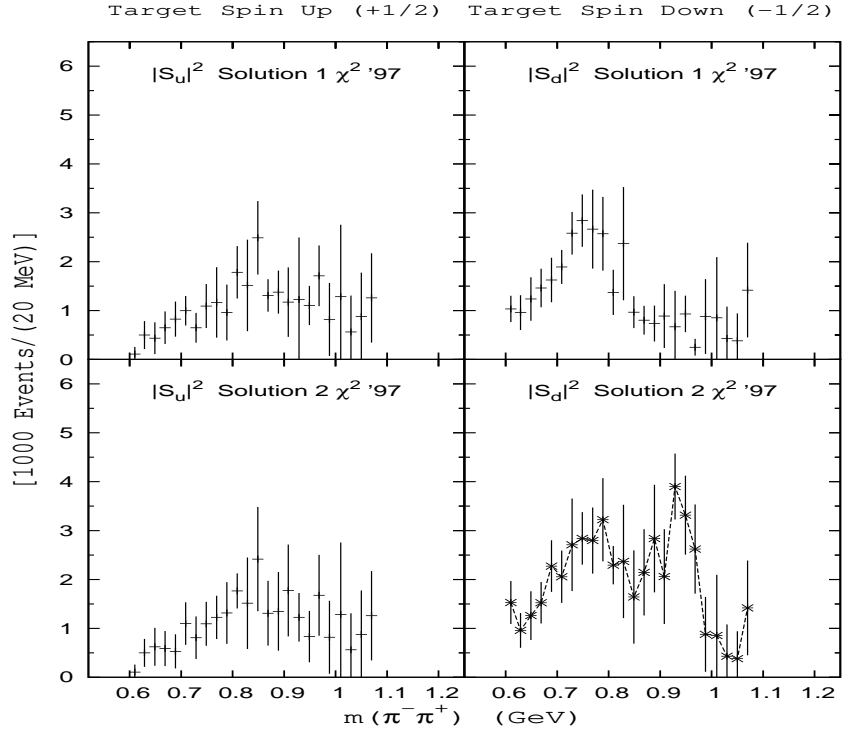}
\caption{$S$-wave moduli $|S_\tau|^2$ in $\pi^- p \to \pi^- \pi^+ n$ at 17.2 GeV/c at low $t$ from analysis $\chi^2$ 97~\cite{kaminski97}.}
\label{Figure 1}
\end{figure}

\begin{figure}[hp]
\includegraphics[width=12cm,height=10.5cm]{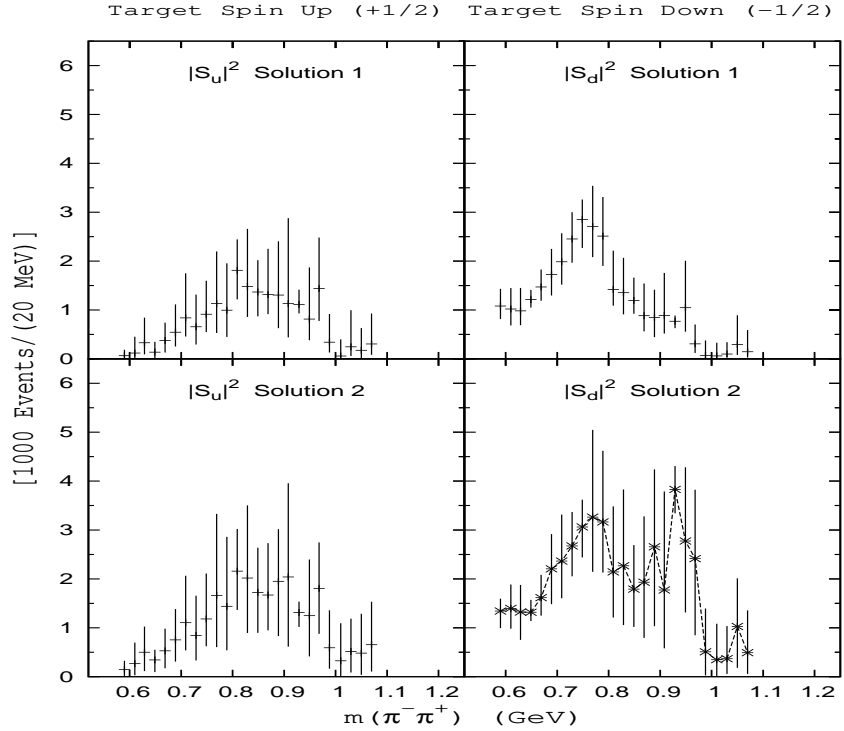}
\caption{Moduli of $S$-wave amplitudes $|S_\tau|^2$ in $\pi^- p \to \pi^- \pi^+ n$ at 17.2 GeV/c at low $t$ from Analysis I~\cite{svec12b}.}
\label{Figure 2}
\end{figure}

\begin{figure} [htp]
\includegraphics[width=12cm,height=10.5cm]{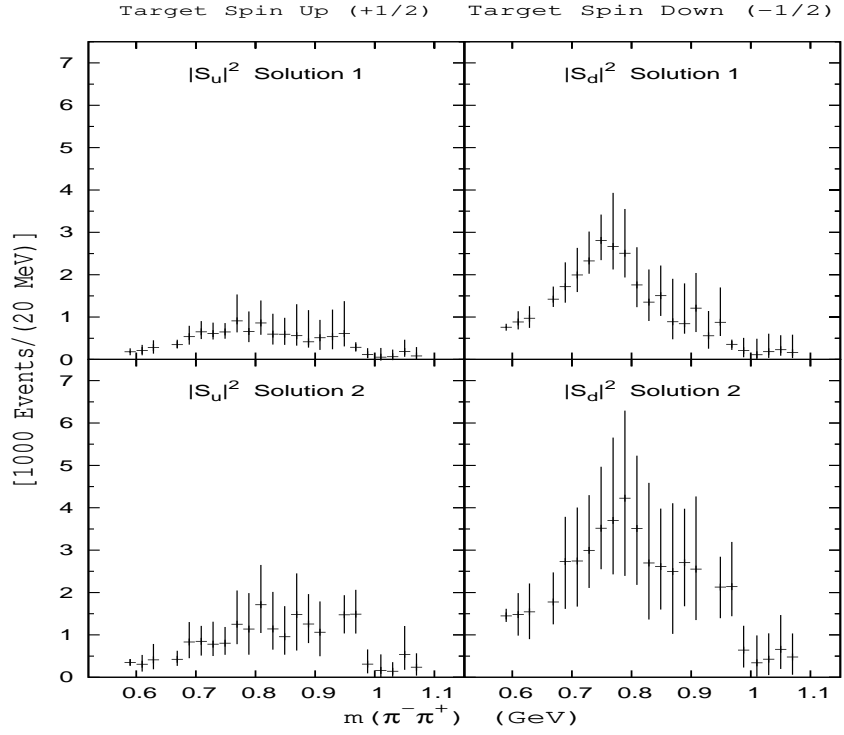}
\caption{The observed $S$-wave spin mixing amplitudes from analysis~\cite{svec14a} using spin mixing mechanism.}
\label{Figure 3}
\end{figure}

\begin{figure} [hp]
\includegraphics[width=12cm,height=10.5cm]{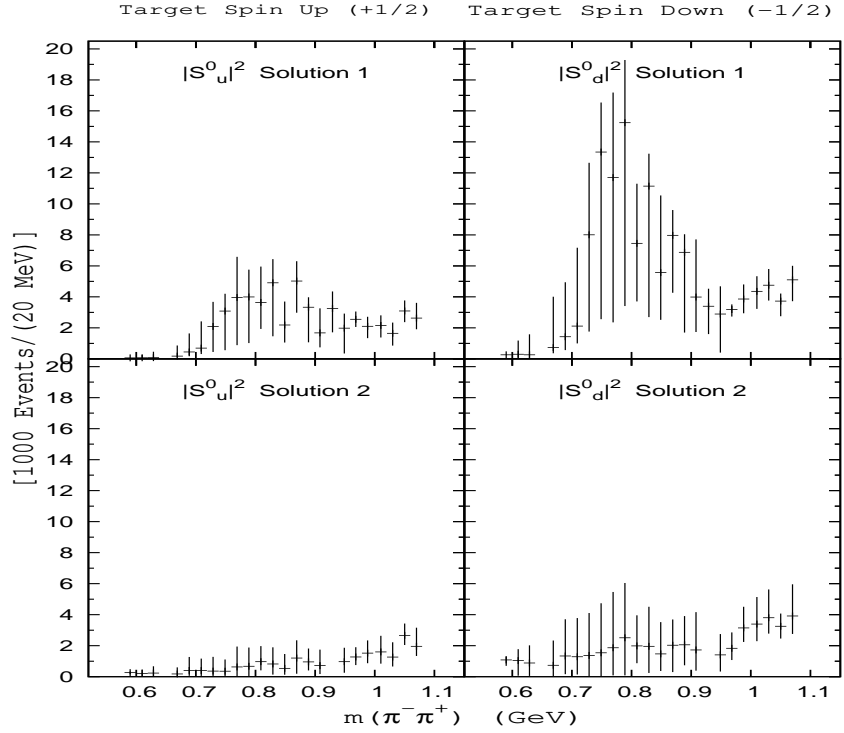}
\caption{The $S$-wave $S$-matrix transversity amplitudes from analysis~\cite{svec14a} using spin mixing mechanism.}
\label{Figure 4}
\end{figure}

\subsection{Helicity amplitudes}

In terms of transversity amplitudes $A^J_{\lambda,\tau}$ with dipion spin $J$ and helicity $\lambda$ the helicity amplitudes $A^J_{\lambda,n}$ are given by~\cite{svec13a}
\begin{equation}
A^J_{\lambda,n}=\frac{(-i)^n}{\sqrt{2}}\bigl(
A^J_{\lambda,u}+(-1)^nA^J_{\lambda,d}\bigr)
\end{equation}
where $n=0,1$ is the helicity flip between the target and recoil nucleons. With the definition (1.3) for the phase $\omega$ and omitting the spin and helicity labels we have for amplitudes $S_n$ and $L_n$
\begin{eqnarray}
|S_n|^2 & = & \bigl(|S_u|^2+|S_d|^2+2(-1)^n|S_u||S_d|\cos\omega\bigr)/2\\
\nonumber
|L_n|^2 & = & \bigl(|L_u|^2+|L_d|^2+2(-1)^n|L_u||L_d|\cos\Omega\bigr)/2
\end{eqnarray}
Here
\begin{equation}
\Omega=\omega + \Phi(L_uS^*_u) - \Phi(L_dS^*_d)
\end{equation}
where $\Phi(L_uS^*_u)$ and $\Phi(L_dS^*_d)$ are the measured relative phases. There is a sign ambiguity in the phases $\Phi(L_uS^*_u)$ and $\Phi(L_dS^*_d)$ with two independent sign assignments $++$ and $+-$ leading to two sets of amplitudes $|L_1|^2++$ and $|L_1|^2+-$. Since the SMM analysis requires $\Phi(L_uS^*_u)=\Phi(L_dS^*_d)>0$ we shall work with the set ++.

There are three analytical solutions for $\omega$~\cite{svec12b}. The only solution which reproduces the resonant shape of $|L_1|^2$ at 770 MeV and satisfies the pion exchange dominance of single flip amplitudes at low $t$ requires $\omega=\pm\pi$. We assume that the phase $\omega$ does not change for the $S$-matrix amplitudes.

Figure 5 shows the helicity amplitudes $|S_1|^2$ and $|L_1|^2$ for the Solutions (1,1) and (2,2) of the transversity amplitudes. The non-flip amplitudes are very small in both Solutions for $S$-and $P$-wave amplitudes. The single flip amplitudes $|L_1|^2$ show the expected $\rho^0(770)$ signal but the $S$-wave amplitudes $|S_1|^2$ also resonate at $\rho^0(770)$ mass indicating $\rho^0(770)-f_0(980)$ spin mixing.

Figure 6 shows the the spin mixing and $S$-matrix helicity amplitudes. We see immediately that the structure at 930 MeV seen in the Solution (2,2) of $|S_d|^2$ is again absent and the $\rho^0(770)$ signal is very clear. Despite large errors, the $S$-matrix amplude $|S^0_1|^2$ is non-resonating below $K\bar{K}$ threshold but rises rapidly above it as expected from the $S$-matrix helicity amplitude. All non-flip amplitudes are very small while the resonating $P$-wave amplitudes $|L^0_1|^2$ dominate at $\rho^0(770)$ mass.  

In the following we reserve the notation (1,1) and (2,2) for the helicity amplitudes from the Standard Analysis I. For the spin mixing and $S$-matrix amplitudes in the SMM amplitude analysis we shall use notation Solution SpinMixing and Solution S-matrix, respectively.

\begin{figure} [htp]
\includegraphics[width=12cm,height=10.5cm]{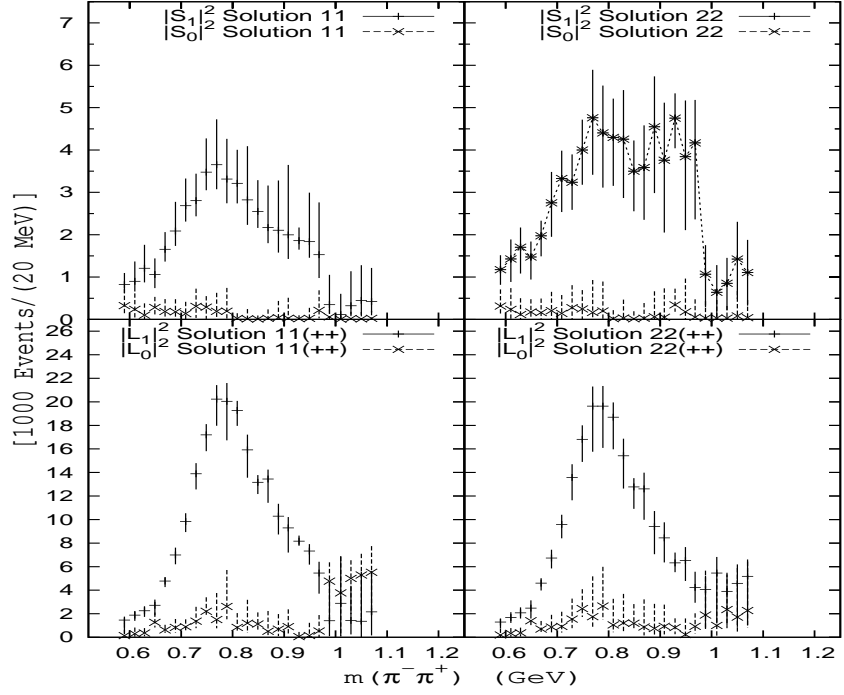}
\caption{Solutions (11) and (22) for helicity amplitudes from Monte Carlo Analysis I~\cite{svec12d}.}
\label{Figure 5}
\end{figure}

\begin{figure} [hp]
\includegraphics[width=12cm,height=10.5cm]{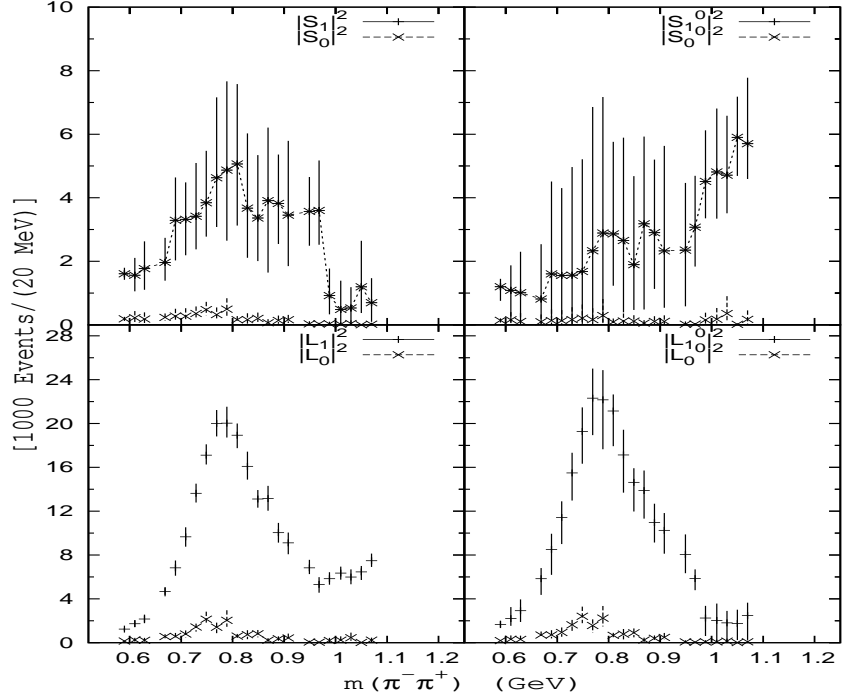}
\caption{Spin mixing and $S$-matrix helicity amplitudes from analysis~\cite{svec14a} using spin mixing mechanism.}
\label{Figure 6}
\end{figure}

\section{$\pi\pi$ phase-shift analysis below the $K\bar{K}$ threshold.}

\subsection{Analytical solutions for the elastic unitarity phase-shift $\delta^0_S$}

The formalism of the $\pi\pi \to \pi\pi $ scattering and its connections to $\pi N \to \pi \pi N$ processes is well known~\cite{martin76,petersen77}. In all previous $\pi\pi$ phase-shift analyses model dependent methods were used to extract the single flip helicity amplitudes from the data which were then related to $\pi \pi$ scattering amplitudes using pion exchange dominance approximation. In our amplitude analysis the helicity amplitudes $|S_1|^2$ and $|L_1|^2$ are model independent and were determined in terms of the measured          transversity amplitudes~\cite{svec12b,svec14a}. In the Section III.C we explain how the linearity of the spin mixing mechanism allows us to use the same form of phase-shift parametrization for the spin mixing and $S$-matrix helicity amplitudes. Despite the spin mixing we treat in the following all amplitudes formally as $S$-matrix analytical amplitudes in accord with all previous analyses and seek a single physical solution for the phase-shifts. 

In our $\pi\pi$ phase-shift analysis we follow closely the method of Estabrooks and Martin~\cite{estabrooks74}. Elementary pion exchange contribution to the single-flip helicity amplitudes $S_1$ and $L_1$ is parametrized by the form
\begin{eqnarray}
S_1 & = & G e^{i\theta_S}C_S \frac{m}{\sqrt{q}}f_S(m) + S_1(NP)\\
\nonumber
L_1 & = &  G e^{i\theta_P}\sqrt{3}\frac{m}{\sqrt{q}}f_P(m) + L_1(NP)
\end{eqnarray}
where for $t$-channel dipion helicity~\cite{estabrooks74}
\begin{equation}
G=N\frac{\sqrt{-t_{av}}}{\mu^2-t_{av}}|F(t_{av}|=
  N\frac{\sqrt{-t_{av}}}{\mu^2-t_{av}}e^{b(t_{av}-\mu^2)}
\end{equation}
is the overall normalization factor at a single value of the momentum transfer $t=t_{av}=0.068$ (GeV/c)$^2$ corresponding to the bin $0.005 < |t| <0.20$ (GeV/c)$^2$, $\mu$ is the pion mass and $m$ and $q=0.5\sqrt{m^2-4\mu^2}$ are the dipion mass and cms momentum, respectively. The vertex factor $C_S$ is introduced in our analysis of the spin mixing amplitudes to normalize $|f_S|^2$ to the value of $|f_S|^2$ at $m=769$ MeV from the KLR 97 analysis corresponding to $\delta^0_S=89.50^\circ$. The factor $\sqrt{3}=\sqrt{2J+1}$ for $J=1$. The terms $S_1(NP)$ and $L_1(NP)$ in (3.1) are the non-pole terms of the amplitudes $S_1$ and $L_1$, repectively. The phases of the pole terms in general differ from the phases of the entire amplitudes $S_1$ and $L_1$
\begin{eqnarray}
\Phi(S_1) & \neq & \Phi(S_1(\text{pole}))=\theta_S+\Phi(f_S)\\
\nonumber
\Phi(L_1) & \neq & \Phi(L_1(\text{pole}))=\theta_P+\Phi(f_P) 
\end{eqnarray}
In the following the non-pole terms $S_1(NP)$ and $L_1(NP)$ will be neglected.

In terms of $\pi\pi$ scattering amplitudes $f^I_L$ with definite isospin $I$ the amplitudes $f_S$ and $f_P$ read
\begin{eqnarray}
f_S & = & \frac{2}{3}f^0_S+\frac{1}{3}f^2_S\\
\nonumber
f_P & = & f^1_P
\end{eqnarray}
Following the Estabrooks-Martin analysis we assume elastic $\pi^- \pi^+$ scattering below $K\bar{K}$ threshold
\begin{equation}
f^I_L=\sin \delta^I_L e^{i\delta^I_L}
\end{equation}
We determine the normalization factor $G$ from the condition that $\delta^1_P=90^\circ$ at the peak value $|L_1^*|^2$ of $|L_1|^2$ at 769 MeV
\begin{equation}
G^2=\frac{1}{3}|L_1^*|^2q^*/m^{*2}
\end{equation}
Then the $P$-wave amplitude reads
\begin{equation}
|f_P|^2= \frac{q}{q^*} \frac{m^{*2}}{m^2} \frac{|L_1|^2}{|L_1^*|^2}=\sin^2 \delta^1_P 
\end{equation}
The $S$-wave amplitude is given by
\begin{equation}
|f_S|^2=\frac{q}{m^2}\frac{|S_1|^2}{G^2C_S^2}
=\frac{4}{9}|f^0_S|^2+\frac{1}{9}|f^2_S|^2+\frac{4}{9}|f^0_S||f^2_S|
        \cos (\delta^0_S-\delta^2_S)
\end{equation}
The equation (3.8) is a quadratic equation for $\sin^2 \delta^0_S$ with two solutions
\begin{equation}
(\sin^2 \delta^0_S)_{1,2}=\frac{1}{2A} \Bigl( B\pm \sqrt{B^2-AC^2} \Bigr)
\end{equation}
where
\begin{eqnarray}
A & = & 4(1+\sin^2 \delta^2_S)^2 + \sin^2 2\delta^2_S\\
\nonumber
B & = & 2C(1+\sin^2 \delta^2_S) + \sin^2 2\delta^2_S\\
\nonumber
C & = & 9|f_S|^2-\sin^2 \delta^2_S
\end{eqnarray}
For the phase-shifts $\delta^2_S$ we take the values from the Figure 1 of Ref.~\cite{kaminski97}. For $|L_1|^2,|S_1|^2$  we take the average experimental values~\cite{svec12b} and calculate the errors on $\delta^0_S$ using error propagation in function of several variables~\cite{taylor97}. The two solutions for $\delta^0_S$ for input helicity amplitudes (1,1) are labeled (1,1)1 and (1,1)2. Similarly the two solutions for $\delta^0_S$ for input helicity amplitudes (2,2) are labeled (2,2)1 and (2,2)2.

\subsection{Results of the elastic unitarity phase-shift analysis}

Figure 7 compares the four solutions for $\delta^0_S$ from the Standard Analysis I. The Solutions (2,2)2 and (1,1)2 are very steep and somewhat akin to $\delta^1_P$ and are rejected. Similarly the Solution (1,1)1 is rejected in favour of the less steep and unique Solution (2,2)1. All four solutions pass through 90$^\circ$ at or near 770 MeV which hints at the presence of the $\rho^0(770)$ resonance in the amplitude $f_S$.

Figure 8 presents the solutions from the SMM Analysis. Both Solutions SpinMixing 1 and SpinMixing 2  pass through 90$^\circ$ at ~770 MeV and are similar to the Solutions (2,2)1 and (2,2)2, respectively. The two Solutions S-Matrix 1 and S-matrix 2 are nearly equal, small and show no hint of $\rho^0(770)$ signal. 

Figures 9 and 10 compare Solutions (2,2)1 and SpinMixing 1 with the Solution EM 74 ($t$-channel) from Estabrooks-Martin analysis on unpolarized target~\cite{estabrooks74}, with the Solution Down-flat from the 1997 analysis KLR 97~\cite{kaminski97} and with the modified Solution Down-flat from the 2002 analysis KLR 02~\cite{kaminski02} which used also $\pi^-p \to \pi^0 \pi^0 n$ data~\cite{gunter01,pi0pi0pwa}. The transversity amplitudes obtained in the $\chi^2$ 97 amplitude analysis~\cite{kaminski97} were used in both phase shift analyses KLR 97 and KLR 02. Solutions (2,2)1, SpinMixing 1 and KLR 97 pass through 90$^\circ$ at 770 MeV. Solutions EM 74 and KLR 02 pass through 90$^\circ$ near 770 MeV. Below 770 MeV all Solutions are similar. Solution (2,2)1 and SpinMixing 1 flaten out at 110$^\circ$-120$^\circ$ between 790-970 MeV. Solutions KLR 97 and KLR 02 flatten out at ~110$^\circ$ between 790-910 MeV but rise to 120$^\circ$ and 130$^\circ$ above 930 MeV, respectively.

We conclude that despite the diverse assumptions about the input helicity amplitudes and different methods of analyses, all solutions are broadly consistent with each other. Within errors our Solutions (2,2)1 and SpinMixing 1 are in a remarkable agreement with the Solution Down-flat KLR 97. 

\begin{figure} [htp]
\includegraphics[width=12cm,height=10.5cm]{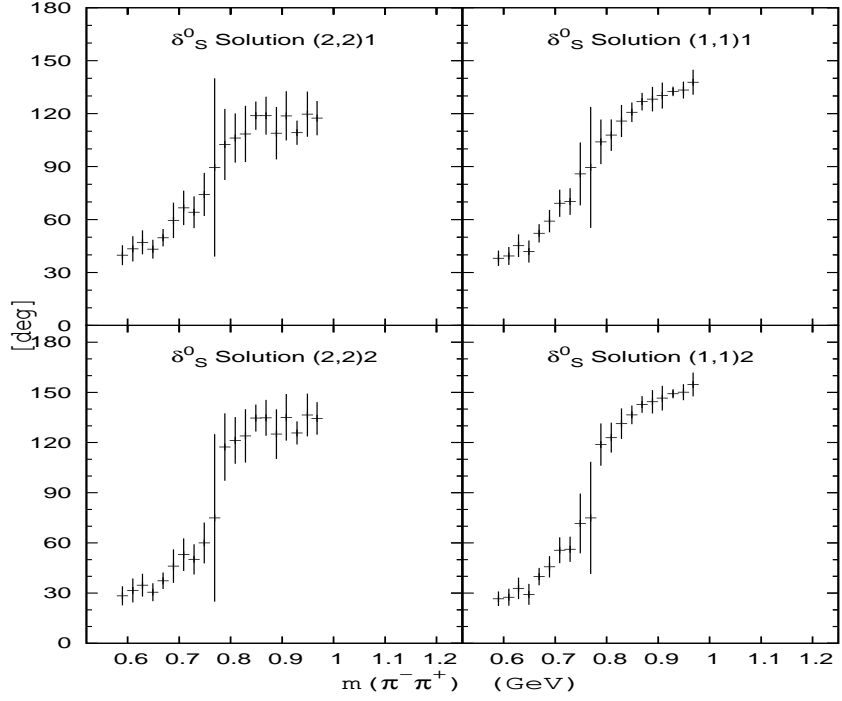}
\caption{Elastic Solutions for the phase shift $\delta^0_S$ from Analysis I below $K\bar{K}$ threshold.}
\label{Figure 7}
\end{figure}

\begin{figure}[hp]
\includegraphics[width=12cm,height=10.5cm]{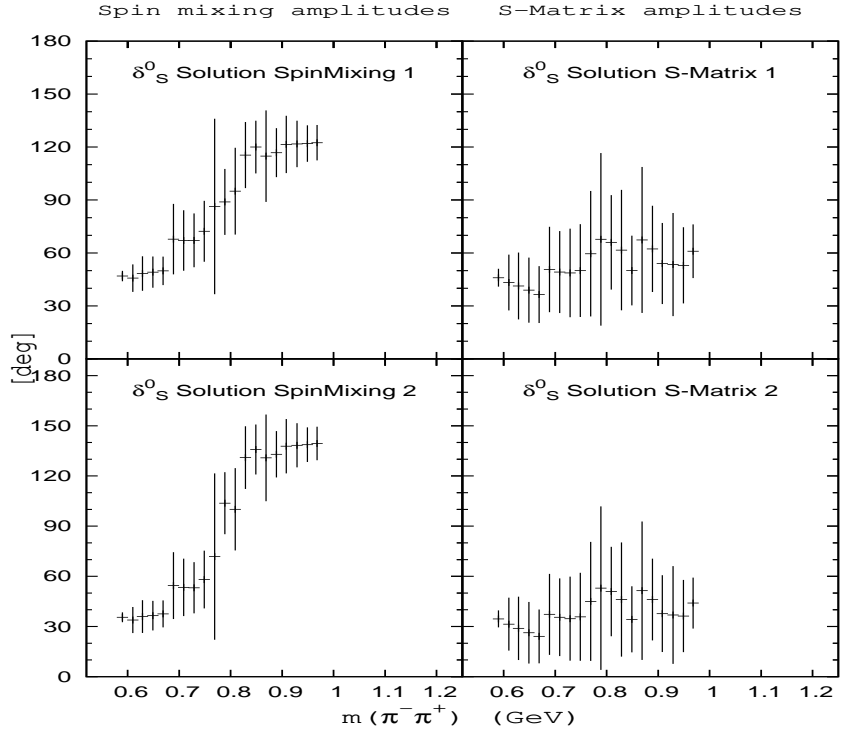}
\caption{Elastic Solutions for $\delta^0_S$ from analysis using spin mixing mechanism below $K\bar{K}$ threshold.}
\label{Figure 8}
\end{figure}

\subsection{A note on $\pi\pi$ phase-shift analysis in the presence of spin mixing}

In Standard Model there is no spin mixing interaction. Since the $S$-and $P$-wave amplitudes $S_\tau, L_\tau, \tau=u,d$ in $\pi^- p \to \pi^- \pi^+ n$ mix spins they are not $S$-matrix amplitudes. They are related to $S$-matrix amplitudes $S_\tau^0, L_\tau^0, \tau=u,d$ by the spin mixing mechanism developed in Ref.~\cite{svec13b}

\begin{figure} [htp]
\includegraphics[width=12cm,height=10.5cm]{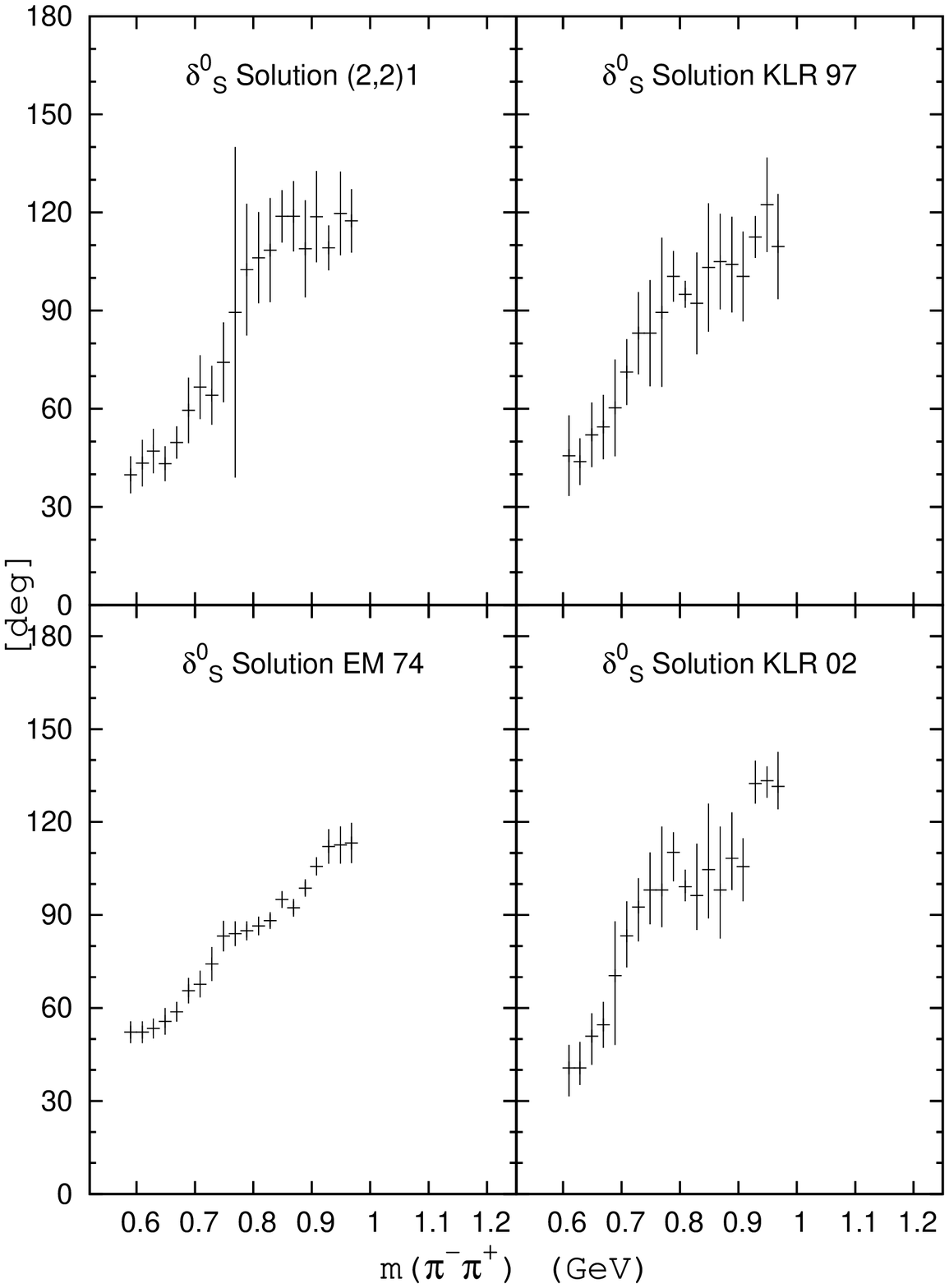}
\caption{Solutions (2,2)1 and EM 74~\cite{estabrooks74} compared with down-flat Solutions KLR 97~\cite{kaminski97} and KLR 02~\cite{kaminski02}.}
\label{Figure 9}
\end{figure}

\begin{figure}[hp]
\includegraphics[width=12cm,height=10.5cm]{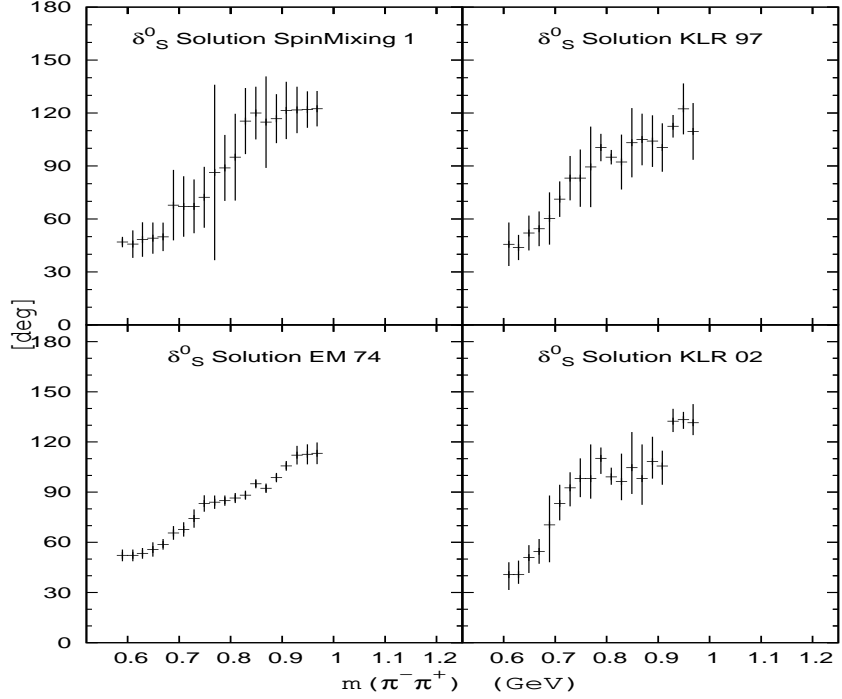}
\caption{Sol. SpinMixing 1 compared with Solutions EM 74~\cite{estabrooks74} and down-flat KLR 97~\cite{kaminski97} and KLR 02~\cite{kaminski02}.}
\label{Figure 10}
\end{figure}

\begin{eqnarray}
L_\tau & = & + e^{i\phi} \bigl ( +\cos \theta S_\tau^0 + e^{i\phi} \sin\theta 
L_\tau^0 \bigr )\\
\nonumber
S_\tau & = & +e^{i\phi} \bigl ( -\sin \theta S_\tau^0 + e^{i\phi} \cos\theta 
L_\tau^0 \bigr )\\
\end{eqnarray}

where $\theta$ and $\phi$ are spin mixing parameters. Identical relations hold for the helicity amplitudes $L_n, S_n$ with the replacement of $\tau=u,d \to n=0,1$. The amplitudes $S_n^0$ and $L_n^0$ refer to $S$-matrix helicity amplitudes in $\pi^- p \to \pi^- \pi^+ n$. Assuming the form (3.1) for the $S$-matrix helicity amplitudes we recover the same form for the spin mixing helicity amplitudes with the effective $\pi\pi$ scattering-like amplitudes defined by

\begin{eqnarray}
G^{eff}e^{i\theta^{eff}_P}f^{eff}_P & = & e^{i\phi} \bigl ( +\cos \theta 
GC_S e^{i\theta_S}f_S + e^{i\phi} \sin\theta Ge^{i\theta_P}f_P \bigr )\\
\nonumber
G^{eff}C^{eff}_Se^{i\theta^{eff}_S}f^{eff}_S & = & e^{i\phi} \bigl ( -\sin \theta GC_S e^{i\theta_S}f_S + e^{i\phi} \cos\theta Ge^{i\theta_P}f_P  \bigr )
\end{eqnarray}
where
\begin{eqnarray}
f_S^{eff} & = & \frac{2}{3}f^{0,eff}_S+\frac{1}{3}f^{2,eff}_S\\
\nonumber
f_P^{eff} & = & f^{1,eff}_P
\end{eqnarray}
There is no spin mixing in $\pi^- p \to \pi^0 \pi^0 n$ and in $\pi^+ p \to \pi^+ \pi^+ n$.

\section{Joint $\pi^-\pi^+$ and $\pi^0\pi^0$ $S$-wave phase-shift analysis below 1080 MeV.}

\subsection{Relation between intensities in $\pi^- p \to \pi^- \pi^+ n$ and $\pi^- p \to \pi^0 \pi^0 n$}

In the following $S_n$ and $S_{0,n}$ denote $S$-wave helicity amplitudes with helicity flip $n=0,1$ for the reactions $\pi^- p \to \pi^- \pi^+ n$ and $\pi^- p \to \pi^0 \pi^0 n$, respectively. Recall that the nucleon helicities are defined in the $s$-channel helicity frame (Section II). Then at large energies and small momentum transfers $t$ the single flip amplitudes are dominated by pion exchange while the non-flip amplitudes are dominated by $a_1$ exchange~\cite{martin76}. The single flip amplitudes have a general form
\begin{eqnarray}
S_1  & = & \sqrt{2}C_\pi T(-+) + Q_1\\
\nonumber
S_{0,1} & = & \sqrt{2}C_{0,\pi} T(00) + Q_{0,1}
\end{eqnarray}
where $\sqrt{2}C_\pi T(-+), \sqrt{2}C_{0,\pi} T(00)$ are pion exchange pole terms and $Q_1,Q_{0,1}$ are non-pole terms. $T(-+)$ and $T(00)$ are $J=even$ partial wave amplitudes in $\pi^-\pi^+ \to \pi^-\pi^+$ and $\pi^-\pi^+ \to \pi^0\pi^0$, respectively 
\begin{eqnarray}
T(-+) & = & +\frac{1}{3}T_0+\frac{1}{6}T_2\\
\nonumber
T(00) & = & -\frac{1}{3}T_0+\frac{1}{3}T_2
\end{eqnarray}
where $T_I,I=0,2$ are the $\pi\pi$ isospin amplitudes 
\begin{equation}
T_I=\frac{\eta_I e^{2i\delta_I} -1}{2i}
\end{equation}
The non-flip amplitudes have a general form
\begin{eqnarray}
S_0  & = & \sqrt{2}C_a t(-+) + Q_0\\
\nonumber
S_{0,0} & = & \sqrt{2}C_{0,a} t(00) + Q_{0,0}
\end{eqnarray}
where $\sqrt{2}C_a t(-+), \sqrt{2}C_{0,a} t(00)$ are $a_1$ exchange pole terms and $Q_0,Q_{0,0}$ are non-pole terms. $t(-+)$ and $t(00)$ are $J=even$ partial wave amplitudes in $\pi^- a_1^+ \to \pi^-\pi^+$ and $\pi^- a_1^+ \to \pi^0\pi^0$, respectively 
\begin{eqnarray}
t(-+) & = & +\frac{1}{3}t_0+\frac{1}{6}t_2\\
\nonumber
t(00) & = & -\frac{1}{3}t_0+\frac{1}{3}t_2
\end{eqnarray}
where $t_I,I=0,2$ are the $\pi a_1 \to \pi\pi$ isospin amplitudes. 

Amplitude analyses of the CERN measurements of $\pi^- p \to \pi^-\pi^+ n$ on polarized target determine transversity production amplitudes of definite $t$-channel naturality. These transversity amplitudes $A_\tau$ with transversity 
$\tau=u,d$ are related to the helicity amplitudes of definite $t$-channel naturality $A_n,n=0,1$ by relations~\cite{svec13a}
\begin{eqnarray}
A_u & = & \frac{1}{\sqrt{2}}(A_0+iA_1)\\
\nonumber
A_d & = & \frac{1}{\sqrt{2}}(A_0-iA_1)
\end{eqnarray}
With these relations the $S$-wave transversity amplitudes in $\pi^- p \to \pi^-\pi^+ n$ read
\begin{eqnarray}
g \equiv S_u & = & C_\pi T(-+)+iC_a t(-+) + R\\
\nonumber
h \equiv S_d & = & C_\pi T(-+)-iC_a t(-+) + \bar{R}
\end{eqnarray}
where the non-pole terms $R=(Q_0+iQ_1)/\sqrt{2}$ and $\bar{R}=(Q_0-iQ_1)/\sqrt{2}$. The $S$-wave transversity amplitudes in $\pi^- p \to \pi^0\pi^0 n$ read
\begin{eqnarray}
g_0 \equiv S_{0,u} & = & C_{0,\pi} T(00)+iC_{0,a} t(00) + R_0\\
\nonumber
h_0 \equiv S_{0,d} & = & C_{0,\pi} T(00)-iC_{0,a} t(00) + \bar{R_0}
\end{eqnarray}
where the non-pole terms $R_0=(Q_{0,0}+iQ_{0,1})/\sqrt{2}$ and $\bar{R_0}=(Q_{0,0}-iQ_{0,1})/\sqrt{2}$. After some simple algebra the amplitudes $g_0$ and $h_0$ can be expressed in terms of the amplitudes $g$ and $h$ as follows
\begin{eqnarray}
g_0 & = & \frac{1}{2}\Bigl(\frac{C_{0,a}}{C_a}+\frac{C_{0,\pi}}{C_\pi} \Bigr)(R-g)
         +\frac{1}{2}\Bigl(\frac{C_{0,a}}{C_a}-\frac{C_{0,\pi}}{C_\pi} \Bigr)(\bar{R}-h)\\
\nonumber
    &   & +\frac{1}{2}C_{0,a} t_2+i\frac{1}{2}C_{0,\pi} T_2 +R_0\\
h_0 & = & \frac{1}{2}\Bigl(\frac{C_{0,a}}{C_a}-\frac{C_{0,\pi}}{C_\pi} \Bigr)(R-g)
         +\frac{1}{2}\Bigl(\frac{C_{0,a}}{C_a}+\frac{C_{0,\pi}}{C_\pi} \Bigr)(\bar{R}-h)\\
\nonumber
    &   & +\frac{1}{2}C_{0,a} t_2-i\frac{1}{2}C_{0,\pi} T_2 +\bar{R_0}    
\end{eqnarray}
To simplify we assume $t_2=0$ and 
\begin{equation}
\frac{C_{0,a}}{C_a}=\frac{C_{0,\pi}}{C_\pi}=K
\end{equation}
Then the equations (4.9) and (4.10) read
\begin{eqnarray}
g_0 & = & K \bigl(+\frac{1}{2}iC_\pi T_2-g\bigr) +R_0+KR\\
\nonumber
h_0 & = & K \bigl(-\frac{1}{2}iC_\pi T_2-h\bigr) +\bar{R_0}+K\bar{R}
\end{eqnarray}
Inverting the relations (4.6) to express helicity amplitudes in terms of transversity amplitudes 
\begin{eqnarray}
A_0 & = & \frac{+1}{\sqrt{2}}(A_u+A_d)\\
\nonumber
A_1 & = & \frac{-i}{\sqrt{2}}(A_u-A_d)
\end{eqnarray}
we find
\begin{eqnarray}
S_{0,1} & = & K \bigl( \frac{1}{\sqrt{2}}C_\pi T_2-S_1 \bigr) +Q_{0,1} +K Q_1\\
\nonumber
S_{0,0} & = & -KS_0 +Q_{0,0} +K Q_0
\end{eqnarray}
Assuming we can neglect the non-pole terms in (4.14) we obtain relations
\begin{eqnarray}
|S_{0,1}|^2 & = & |K|^2 \bigl(|S_1|^2 + \frac{1}{2}|C_\pi T_2|^2 -
\sqrt{2}Re(C_\pi T_2 S^*_1)\bigr)\\
\nonumber
|S_{0,0}|^2 & = & |K|^2 |S_0|^2
\end{eqnarray}
Following KLR 02~\cite{kaminski02} we assume that the factors $C_\pi$ and $C_{0,\pi}$ differ only in phase so that the modulus $|K|^2=1$. Then we obtain the final relation
\begin{equation}
I_0= I + \frac{1}{2}|C_\pi T_2|^2 -\sqrt{2}Re(C_\pi T_2 S^*_1)
\end{equation}
where $I_0=|S_{0,0}|^2+|S_{0,1}|^2$ and $I=|S_0|^2+|S_1|^2$ are the measured $S$-wave intensities in $\pi^- p \to \pi^0 \pi^0 n$~\cite{gunter01,pi0pi0pwa} and $\pi^- p \to \pi^- \pi^+ n$~\cite{kaminski97,svec12b,svec14a}, respectively. The amplitude $S_1$ in (4.16) is the exact single flip $S$-wave amplitude in $\pi^- p \to \pi^- \pi^+ n$.

In their paper~\cite{kaminski02} the authors provide more detailed relations between the transversity amplitudes in the $\pi^0\pi^0$ and $\pi^-\pi^+$ channels which read
\begin{eqnarray}
g_0 & = & a_g T_2 + b_g g + b_h h\\
\nonumber
h_0 & = & a_h T_2 + b_h^* g + b_g^* h
\end{eqnarray}
where $a_g,a_h,b_g,b_h$ are kinematical factors. These relations involve non-leading terms and reflect the difference in the binning in momentum transfers $|t|<0.20$(GeV/c)$^2$ with $\Delta t=0.1900$(GeV/c)$^2$ and $\Delta t_1=0.1950$(GeV/c)$^2$ in the $\pi^0 \pi^0$ and $\pi^- \pi^+$ channels, respectively. From (4.17) we find 
\begin{eqnarray}
I_0 & = & (|b_g|^2+|b_h|^2)I+(|a_g|^2+|a_h|^2)|T_2|^2+4 Re(b_gb_h^* gh^*)\\
\nonumber
    &   & +2 Re[(a_gb_g^*+a_hb_h)T_2g^*] + 2 Re[(a_gb_h^*+a_hb_g)T_2h^*]
\end{eqnarray}    
With the approximations
\begin{equation}
\frac{\Delta t}{\Delta t_1} \sim 1, \quad \sin \Theta_s \sim 0
\end{equation}
where $\Theta_s$ is the neutron scattering angle with respect to proton, we find
\begin{eqnarray}
a_g \to +\frac{1}{2} C_\pi, & \quad & a_h \to -\frac{1}{2} C_\pi\\
\nonumber
b_g \to -1, & \quad & b_h \to 0
\end{eqnarray}
and we recover the relation (4.16).

\subsection{Analytical solutions for the joint $\pi\pi$ phase-shift analysis} 

The explicit form of the amplitude $S_1$ using (4.13) reads
\begin{equation}
S_1=\frac{-i}{\sqrt{2}}e^{i\Phi(S_u)}(|S_u|-|S_d|e^{i\omega})
\end{equation}
The vertex factor $C_\pi=|C_\pi|\exp{i\theta_S}$ and $T_2=\sin \delta^0_S \exp{i\delta^0_S}=-|T_2|\exp{i\delta^0_S}$ since $\delta^0_S<0$. Then the last term in (4.16) takes the form
\begin{equation}
-\sqrt{2}Re(C_\pi T_2 S^*_1)=|C_\pi|| T_2||S_u|\cos\theta - |C_\pi|| T_2||S_d|\cos(\theta -\omega)
\end{equation}
where 
\begin{equation}
\theta=\theta_S+\Phi(T_2)+\pi/2-\Phi(S_u)
\end{equation}
In our analysis $\omega=\pm\pi$ and the equation (4.16) reads
\begin{equation}
I_0= I + \frac{1}{2}|C_\pi T_2|^2 +\sqrt{2}|C_\pi|| T_2||S_1|\cos\theta 
\end{equation}
where $|S_1|=\frac{1}{\sqrt{2}}(|S_u|+|S_d|)$ and $|T_2|=|\sin \delta^2_S|$ are known. We rewrite (4.23) to define a new phase $\chi$ 
\begin{equation}
\chi=\Phi(T_2)-\theta=-\theta_S-\pi/2+\Phi(S_u)
\end{equation}
Then the equation (3.1) for $S_1$ can be written in the form 
\begin{equation}
e^{i\chi} |S_1|= \frac{1}{\sqrt{2}} |C_\pi| f_S
\end{equation}
or in an equivalent form
\begin{equation}
e^{i\chi} |f_S|=f_S
\end{equation}
If the vertex factor $|C_\pi|$ is known then we can use (4.24) and (4.26) to determine $\theta$ and $|f_S|$, respectively. With $\chi$ and $|f_S|$ known the real and imaginary parts of the equation (4.27) represent two equations from which we can determine analytically a unique solution for the phase shift $\delta^0_S$ and the inelasticity $\eta^0_S$ provided we assume that $\eta^2_S=1$ and $\eta^1_P=1$. Using the definitions (3.4) and (4.3) we then find from (4.27)
\begin{eqnarray}
A=+\eta^0_S\sin 2\delta^0_S & = & 3\cos \chi |f_S| -\frac{1}{2} \sin 2\delta^2_S>0\\
\nonumber
B=-\eta^0_S\cos 2\delta^0_S & = & 3\sin \chi |f_S| -\sin^2 \delta^2_S -1
\end{eqnarray}
Then the solutions for $0<\delta^0_S$ and $0<\eta^0_S$ are given by
\begin{eqnarray}
\tan 2\delta^0_S = \frac{A}{-B}\\
\eta^0_S = \sqrt{A^2+B^2}
\end{eqnarray}

The vertex factor $|C_\pi|$ is related to the vertex correction factor $C_S$ 
\begin{equation}
|C_\pi|=\sqrt{2}GC_S\frac{m}{\sqrt{q}}
\end{equation}
As an initial step we used for each set of input helicity amplitudes the vertex correction factors $C_S$ determined by the elastic analyses to calculate $\cos \theta$ and $|f_S|$. For $\theta >0$ we obtained very large inelasticities and unreasonable phase shifts. For $\theta<0$ we obtaine large inelasticities only at low masses for Solutions (2,2) joint and SpinMixing joint while no physical solutione were found below 680 MeV for Solutions (1,1) joint and $S$-Matrix joint. A guesswork was required to adjust $C_S$ to obtain more satisfactory solutions.

To avoid the guesswork we need an independent auxiliary equation to estimate better the factors $C_S$ for each input helicity amplitudes at each mass bin. We use $\chi=\delta^0_S-\theta$ and (4.26) to write $\eta^{02}_S = A^2+B^2$ in the form
\begin{eqnarray}
|C_\pi|^2 \eta^{02}_S & = & 18|S_1|^2+12\Bigl(\sqrt{2}|C_\pi||T_2||S_1|\cos \theta \Bigr)\\
\nonumber
   &  + &6\sqrt{2}|C_\pi|\cos\delta^2_S|S_1|\sin\theta + |C_\pi|^2\Bigl( 1+5\sin^2\delta^2_S\Bigr)
\end{eqnarray}
From (4.24) we have
\begin{equation}
\sqrt{2}|C_\pi||S_1|\cos\theta=|C_\pi|^2 C_1 +C_2
\end{equation}
where
\begin{eqnarray}
C_1 & = & \frac{1}{2}\sin \delta^2_S\\
\nonumber
C_2 & = & -\frac{I_0-I}{\sin\delta^2_S}
\end{eqnarray}
Substituting (4.33) into (4.32) we can write
\begin{equation}
-\sqrt{2}|C_\pi||S_1|\sin\theta  = |C_\pi|^2 C_3 + C_4
\end{equation} 
where
\begin{eqnarray}
C_3 & = & \frac{1-\sin^2\delta^2_S-\eta^{02}_S}{6\cos\delta^2_S}\\
\nonumber
C_4 & = & \frac{3|S_1|^2+2(I_0-I)}{\cos\delta^2_S}
\end{eqnarray}
Taking a square of the equations (4.33) an (4.35) and adding them we get a quadratic equation for $X=|C_\pi|^2$
\begin{equation}
X^2A-2XB+C=0 
\end{equation}
where
\begin{eqnarray}
A & = & C_1^2+C_3^2\\
\nonumber
B & = & |S_1|^2-C_1C_2-C_3C_4\\
\nonumber
C & = & C_2^2+C_4^2
\end{eqnarray}
There are two roots $|C_\pi(+)|^2$ and $|C_\pi(-)|^2$ 
\begin{equation}
|C_\pi(\pm)|^2=\frac{1}{A}\{B\pm\sqrt{B^2-AC}\}
\end{equation}
Thus there are two solutions for the vertex correction factor $C_S^2(+)$ and $C_S^2(-)$. 

The equations (4.24), (4.30) and (4.39) constitute a system of three non-linear simultaneous equations for three unknowns $|C_\pi|$, $\theta$ and $\eta^0_S$. Each root $|C_\pi(\pm)|^2$ of (4.37) is a function $h_\pm$ of only $\eta^0_S$. We can write these equations in a general form as two systems of three equations each with different $h_{\pm}$
\begin{eqnarray}
I_0-I & = & f(|C_\pi|,\cos \theta)\\
\eta^0_S & = & g(|C_\pi|,\cos \theta,\sin \theta)\\
|C_\pi|^2 & = & h_{\pm}(\eta^0_S)
\end{eqnarray}
Substituting from (4.41) into (4.42) we obtain
\begin{equation}
|C_\pi|^2 = \tilde{h}_{\pm}(|C_\pi|,\cos \theta,\sin \theta)
\end{equation}
Solving (4.40) and (4.43) for $|C_\pi(\pm)|$ and $\theta(\pm)$ we can calculate $\eta^0_S(\pm)$ and $\delta^0_S(\pm)$. In priciple a unique solution for $\eta^0_S$ and $\delta^0_S$ is possible. However, the equations (4.40) and (4.43) are a highly non-linear system of equations which is extremally difficult to solve. 

To render the system tractable we adopt a more economic approach and consider the relation (4.42) as a generating equation for the vertex factor $|C_\pi(\pm)|^2$ assuming a constant value of the input parameter $\eta^0_S=const$. The phase $\theta(\pm)$ is then determined from (4.40) and the inelasticity $\eta^0_S(\pm)$ and phase-shift $\delta^0_S(\pm)$ can be calculated from (4.30) and (4.29), respectively. Note that the equations (4.40),(4.41) and (4.42) no longer form a system of simultaneous equations for $|C_\pi|^2$ and $\eta^0_S$. Thus the inelasticity $\eta^0_S(\pm)$ calculated from (4.30) will differ from the parameter $\eta^0_S$ in the independent auxiliary equation (4.42) which only serves to provide an estimate of the vertex factor $|C_\pi(\pm)|^2$. To distinguish the two quantities we shall use in the following the notation $\eta^0_S$ for the calculated inelasticity and $\eta=const$ for the independent input parameter in (4.42).

The proposed solution is an approximate but unique analytical solution of the system (4.40)-(4.42). The approximation can be considered acceptable if the calculated inelasticities have physical values below 1 and are not too far from $\eta$. This turns out to be the case.

\subsection{Data on $S$-wave intensities in $\pi^- p \to \pi^0 \pi^0 n$ and $\pi^- p \to \pi^- \pi^+ n$}

In our joint phase-shift analysis of $\pi^-\pi^+$ and $\pi^0\pi^0$ data we shall use for the $\pi^0 \pi^0$ channel the BNL data at 18.3 GeV/c~\cite{gunter01,pi0pi0pwa}. The BNL data were converted from native BNL units "intensity/40 MeV" into our units "1000 events/20 MeV" using a conversion factor $F=0.6700\times10^{-4}$. We obtained this factor by comparing the $f_2(1270)$ peak value in their Figure 5F given in units "intensity/40 MeV" with the value of coresponding 4 bins at $f_2(1270)$ peak in their Figure 4a given in units "events/10 MeV". The data in two bins $0.01 < |t| < 0.10$ (GeV/c)$^2$ and $0.10 < |t| < 0.20$ (GeV/c)$^2$ were combined by addition to a sigle bin $0.01 < |t| < 0.20$ (GeV/c)$^2$ corresponding to the CERN measurements. The data were then interpolated to 20 MeV bins and scaled to 17.2 GeV/c using phase and flux factor $K(s,m^2)$ given by~\cite{svec97a}
\begin{eqnarray}
K(s,m^2) & = & {G(s,m^2) \over {\text{Flux}(s)}}\\
\nonumber
G(s,m^2) & = & {1 \over {(4 \pi )^3}}{q(m^2) \over {\sqrt{[s-(M+\mu)^2][s-(M-\mu)^2]}}}\\
\nonumber
\text{Flux}(s) & = & 4Mp_{\pi lab}
\end{eqnarray}
where $q(m^2)={1 \over {2}} \sqrt{m^2 -4\mu^2}$ is the pion momentum in the center of mass of the dipion system of mass $m$, and $M$ and $\mu$ are the nucleon and pion mass, respectively. The two Solutions 1 and 2 for intensities $I_0=I_S(00)$ are shown in Figure 11 and compared with the corresponding intensities $I=I_S(-+)$ in $\pi^- p \to \pi^- \pi^+ n$ from our Analysis I.

\begin{figure} [htp]
\includegraphics[width=12cm,height=10cm]{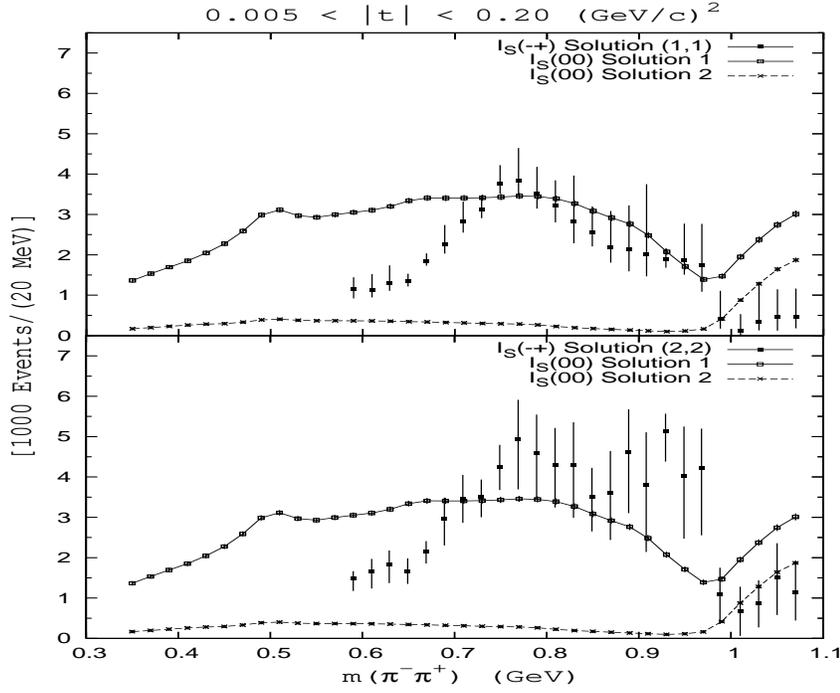}
\caption{Comparison of intensity $I=I_S(-+)$ from Analysis I~\cite{svec12b} with the intensity $I_0=I_S(00)$ from E852~\cite{gunter01}.}
\label{fig11}
\end{figure}

\subsection{Results of the joint phase-shift analysis}

Motivated by the elastic phase-shift analysis we set the generating parameter $\eta=1$ in $C_3$ to determine the two roots $|C_\pi(\pm)|^2$. We used (4.24) and (4.26) to calculate $\cos \theta$ and $|f_S|$, respectively, for each input helicity flip amplitude $|S_1|$ for both roots of the vertex factor. There are positive and negative solutions for $\theta$. There are no physical solutions for $\delta^0_S$ for the vertex factor $|C_\pi(+)|^2$ for either sign of $\theta$. There are no physical solutions for $\delta^0_S$ for the vertex factor $|C_\pi(-)|^2$ with $\theta>0$. Figures 12 and 13 show the vertex correction factors $C_S^2(-)$ calculated from $|C_\pi(-)|^2$ and the corresponding $\theta<0$ for which physical solutions exist for the pairs of input helicity amplitudes (2,2),(1,1) and SpinMixing, S-Matrix, respectively. 

Figures 14 and 15 show the Solutions (2,2) joint and (1,1) joint for the phase-shift $\delta^0_S$ and inelasticity $\eta^0_S$. Comparison of Solution (2,2) joint with the Solution Down-flat KLR $02$ and similar comparison of the Solution (1,1) joint with the Solution Up-flat KLR $02$ shows a remarkable agreement in the phase shift $\delta^0_S$ of the two $\pi\pi$ phase-shift analyses. In both analyses the phase shift $\delta^0_S$ reaches $90^\circ$ near 770 Mev but then it is flat or it drops instead of rising like in the Solutions (2,2)1 and KLR $97$. This appears to be the principal impact of the $\pi^0\pi^0$ data on both phase-shift analyses. The two analyses differ chiefly in the inelasticity. Both Solutions (2,2) joint and (1,1) joint show clearly only physical values $\eta^0_S<1$ in contrast to unphysical values of $\eta^0_S$ in seven mass bins in the analysis KLR $02$.

\begin{figure} [htp]
\includegraphics[width=12cm,height=10.5cm]{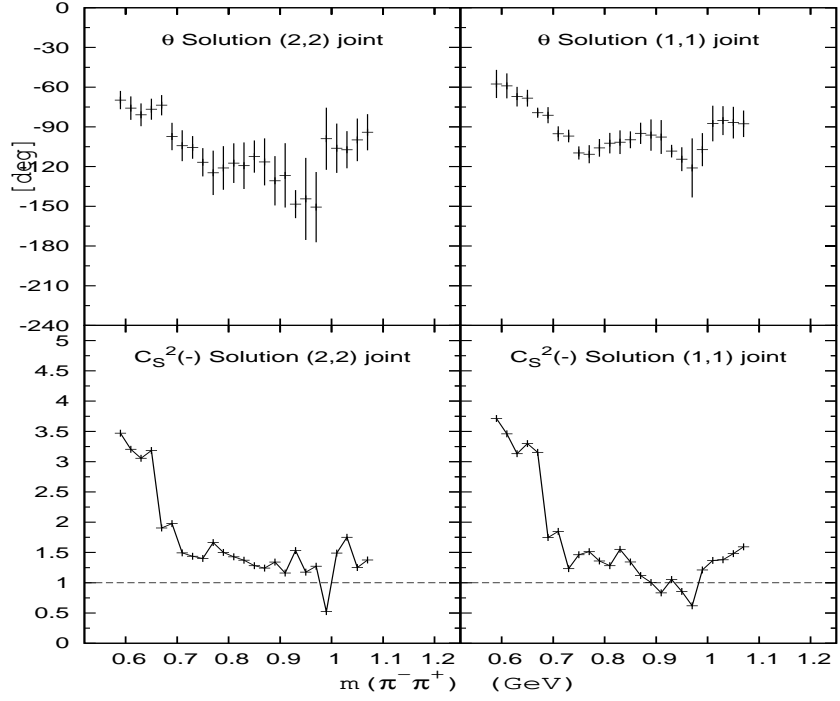}
\caption{Phase $\theta$ and vertex correction factor $C_S^2(-)$ for Solutions (2,2) joint and (1,1) joint.}
\label{Figure 12}
\end{figure}

\begin{figure}[hp]
\includegraphics[width=12cm,height=10.5cm]{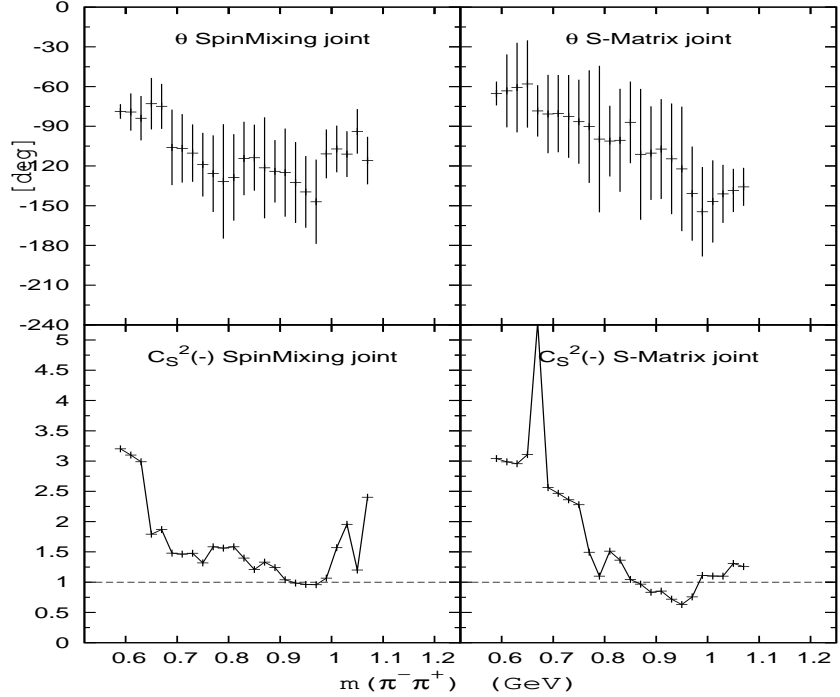}
\caption{Phase $\theta$ and vertex correction factor $C_S^2(-)$ for Solutions SpinMixing joint and S-Matrix joint.}
\label{Figure 13}
\end{figure}

\begin{figure} [htp]
\includegraphics[width=12cm,height=10.5cm]{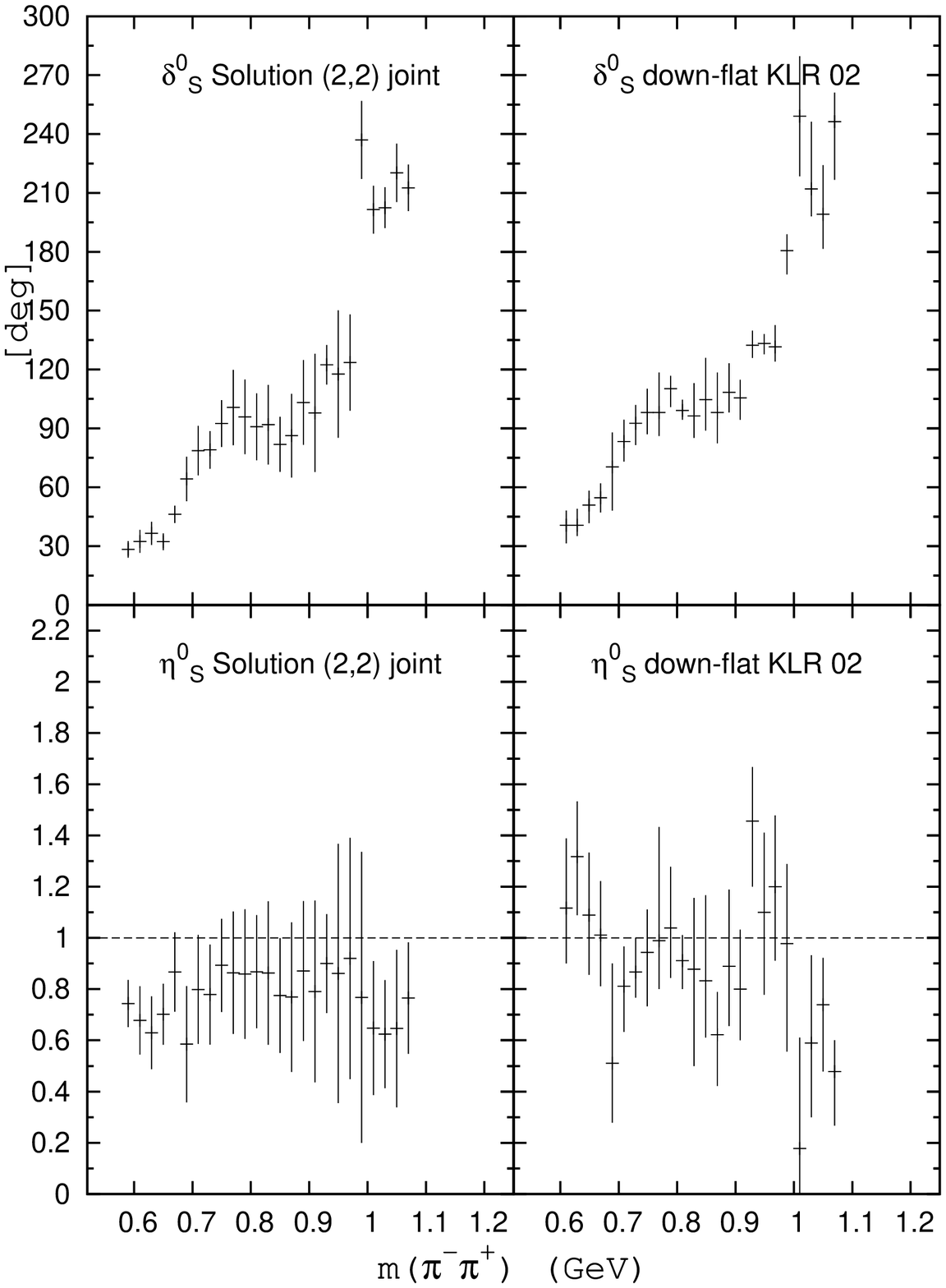}
\caption{Solution (2,2) joint for $\delta^0_S$ and $\eta^0_S$ compared with Solution Down-flat KLR $02$\cite{kaminski02}.}
\label{Figure 14}
\end{figure}

\begin{figure}[hp]
\includegraphics[width=12cm,height=10.5cm]{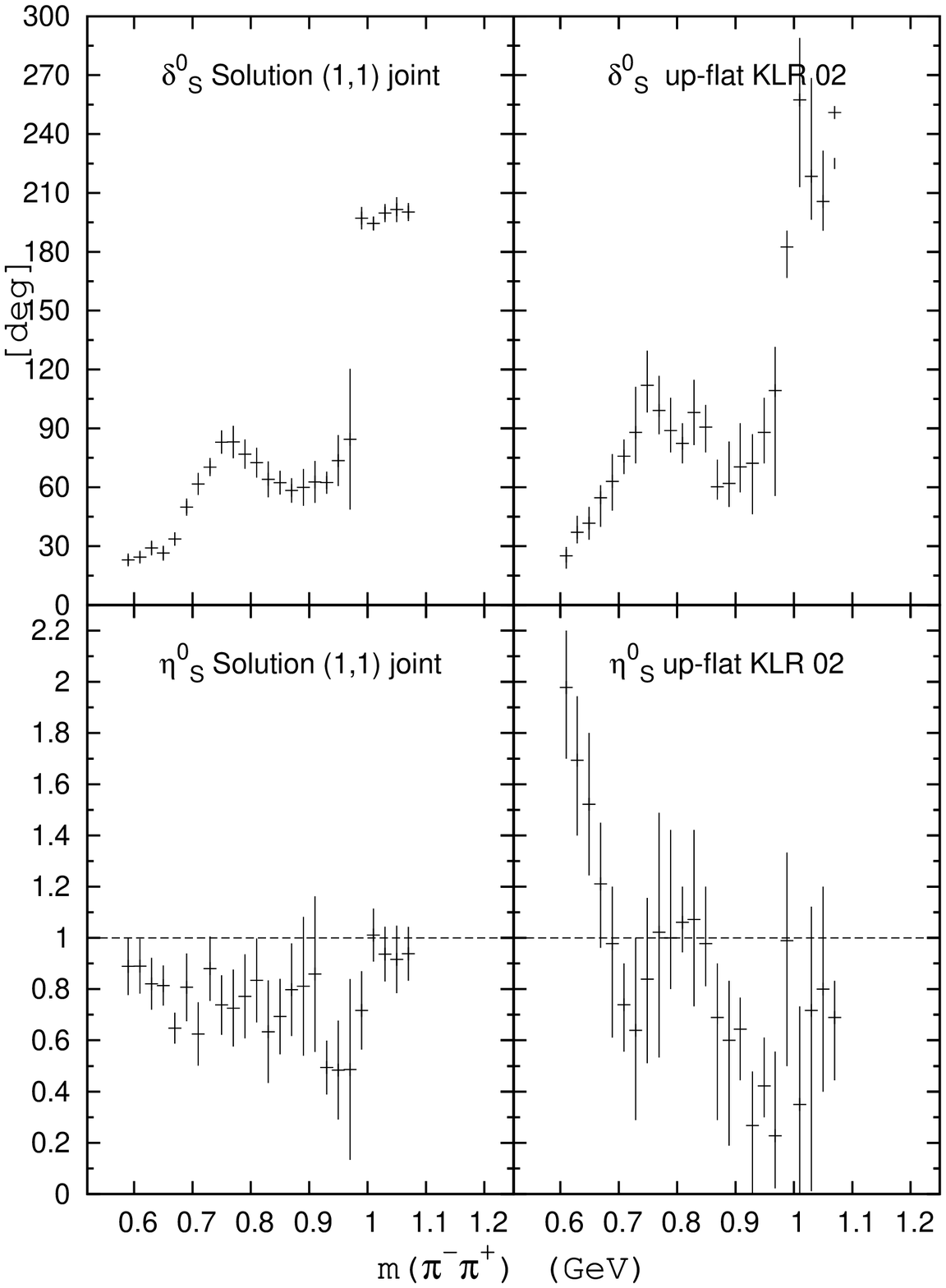}
\caption{Solution (1,1) joint for $\delta^0_S$ and $\eta^0_S$ compared with Solution Up-flat KLR $02$\cite{kaminski02}.}
\label{Figure 15}
\end{figure}

\begin{figure} [htp]
\includegraphics[width=12cm,height=10.5cm]{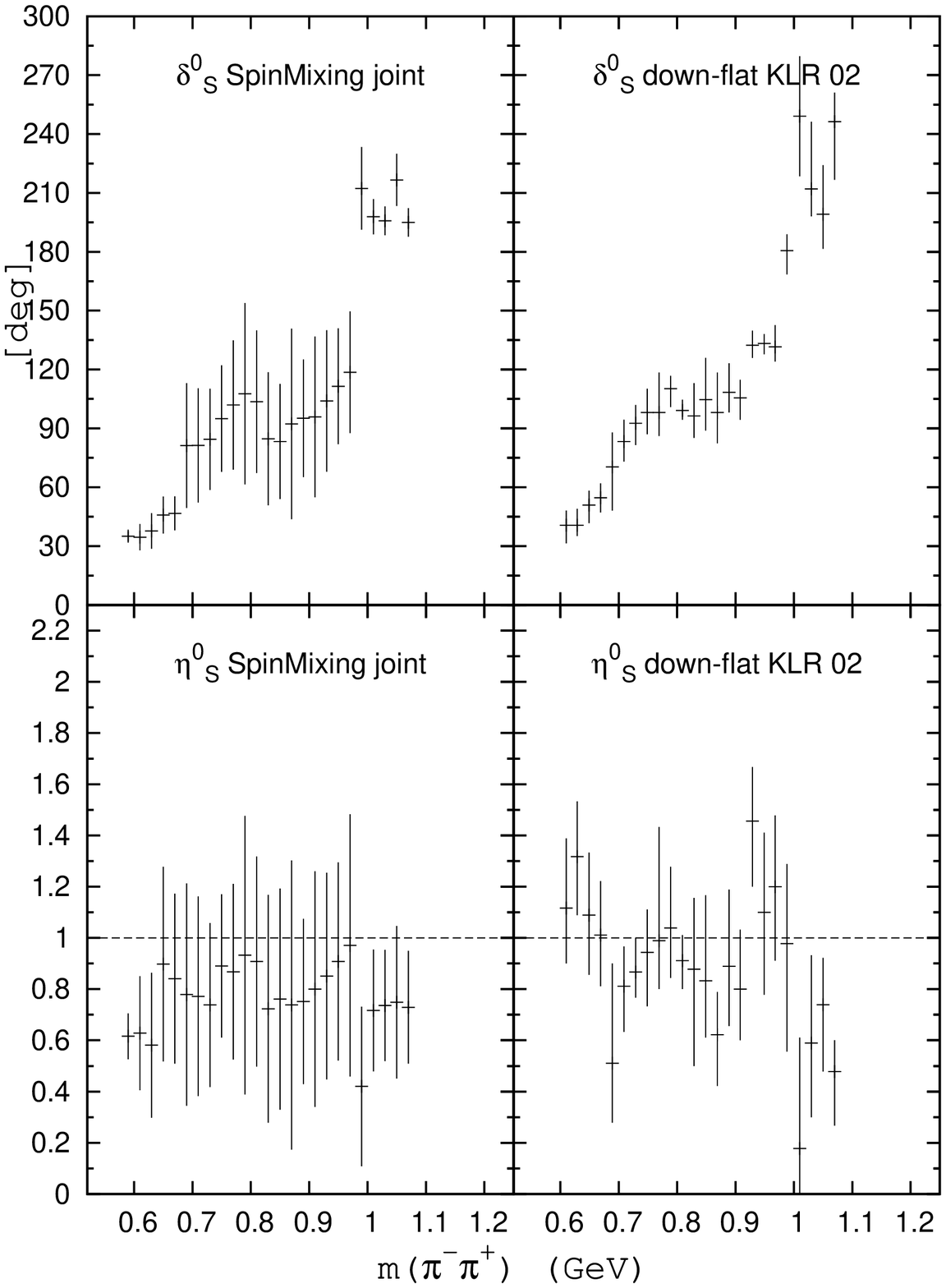}
\caption{Solution SpinMixing joint for $\delta^0_S$ and $\eta^0_S$ compared with Solution Down-flat KLR $02$\cite{kaminski02}.}
\label{Figure 16}
\end{figure}

\begin{figure}[hp]
\includegraphics[width=12cm,height=10.5cm]{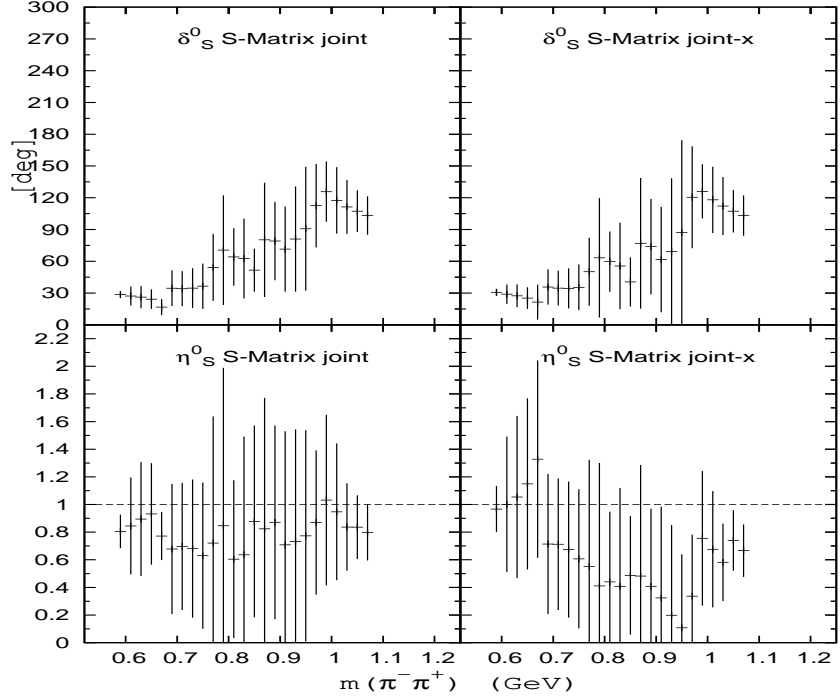}
\caption{Solution S-Matrix joint for $\delta^0_S$ and $\eta^0_S$ compared with Solution S-Matrix joint-x with factor $C_S$(x).}
\label{Figure 17}
\end{figure}

\begin{figure} [htp]
\includegraphics[width=12cm,height=10.5cm]{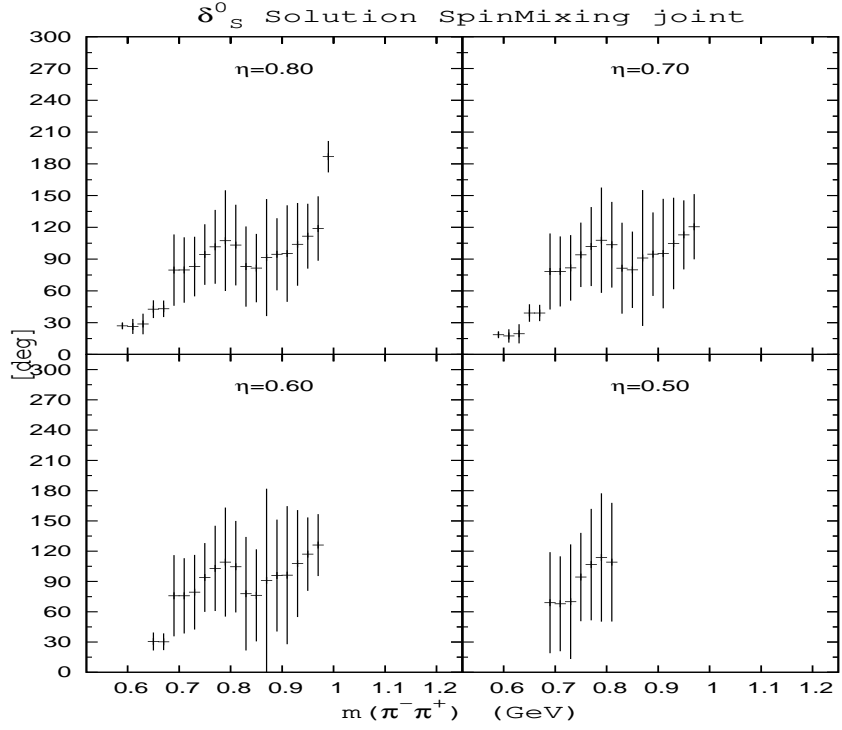}
\caption{Dependence of the phase-shift $\delta^0_S$ on the parameter $\eta$ in Solution SpinMixing joint.}
\label{Figure 18}
\end{figure}

\begin{figure} [htp]
\includegraphics[width=12cm,height=10.5cm]{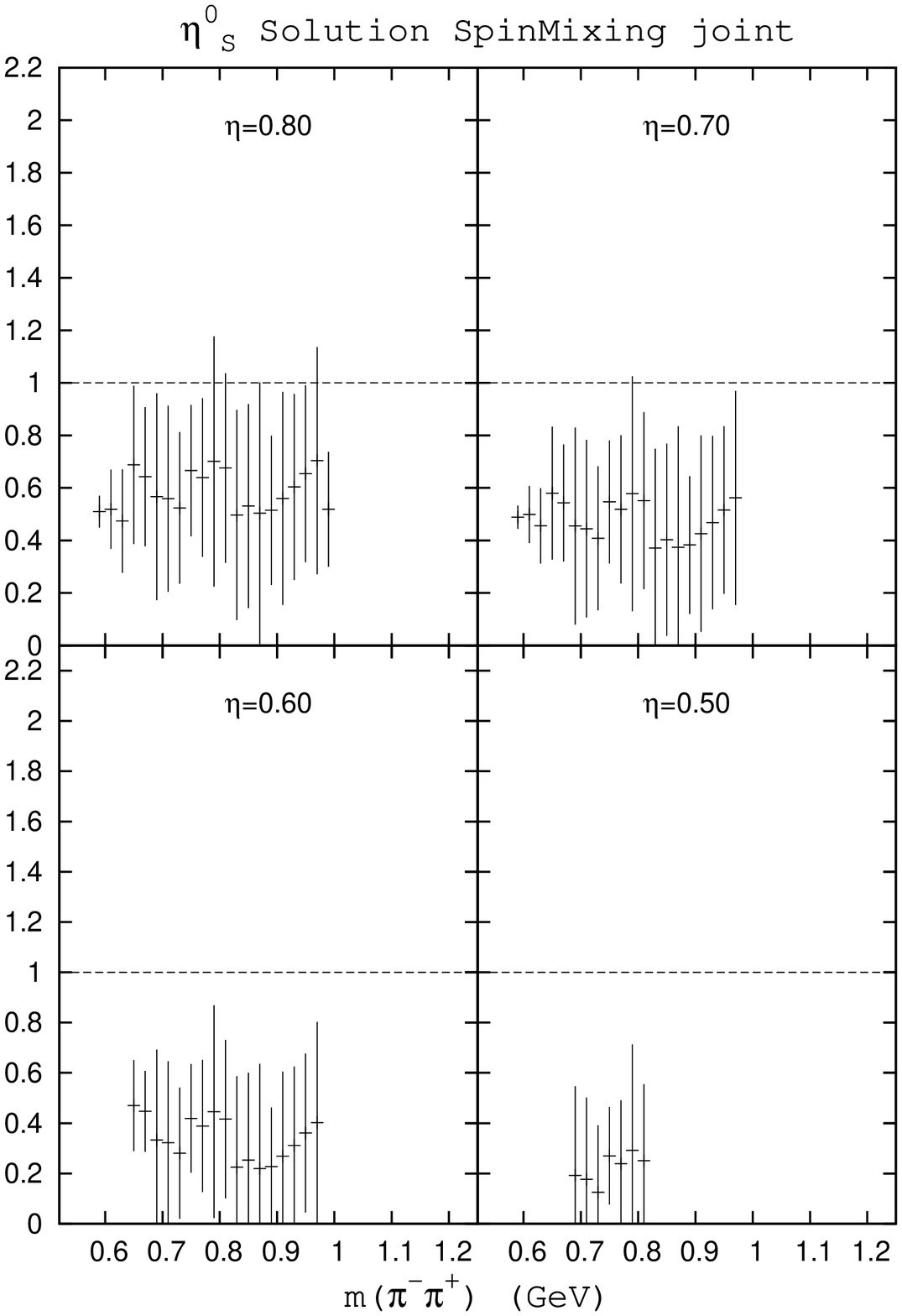}
\caption{Dependence of the inelasticity $\eta^0_S$ on the parameter $\eta$ in Solution SpinMixing joint.}
\label{Figure 19}
\end{figure}

Figure 16 compares the Solution SpinMixing joint with the Solution Down-flat KLR $02$. Despite larger errors there is no suprise in the agreement of $\delta^0_S$ since the helicity amplitudes (2,2) and SpinMixing are similar. The two analyses again differ in the inelasticity with $\eta^0_S<1$ in Solution SpinMixing joint.

Figure 17 shows the result for Solution S-Matrix joint. To illustrate the guesswork solutions this Solution is compared with our best guesswork Solution S-matrix joint-x with the vertex correction 
factor $C_S^2(x)$ given by
\begin{eqnarray}
           m \leq 0.750: &  C_S^2(x) = & C_S^2(1)/0.60\\
\nonumber
0.770 \leq m \leq 0.850: &  C_S^2(x) = & C_S^2(1)/0.75\\
\nonumber
0.870 \leq m \leq 1.070: &  C_S^2(x) = & C_S^2(1)
\end{eqnarray}
where $C_S^2(1)=1.43674$ is the vertex correction factor of the elastic Solution SpinMixing and the mass $m$ is in GeV. With no evidence for $\rho^0(770)-f_0(980)$ mixing both solutions for $\delta^0_S$ are slowly rising. 
The two analyses differ in the inelasticity with $\eta^0_S<1$ in Solution SpinMixing joint while $\eta^0_S>1$ at low masses in the other Solution.

We have also examined the dependence of the Solutions SpinMixing joint and S-Matrix joint on the parameter $\eta$ by repeating the calculations for values of $\eta=0.90-0.50$ in steps of 0.10. The values of $\delta^0_S$ do not change much but the inelasticity $\eta^0_S$ is decreasing. There is an increase of mass bins with no physical solutions with decreasig $\eta$ due to $B^2-AC<0$ in (4.39) which becomes very rapid for $\eta \leq 0.60$. At $\eta=0.50$ in Solution SpinMixing joint there are no physical solutions for $m \leq$ 670 MeV and for $m \geq$ 830 MeV. In the Solution S-Matrix joint at this value of $\eta$ there are no physical solutions at all except at a single mass bin $m=$790 MeV. The situation is visualized for the Solution SpinMixing in the Figures 18 and 19 where we notice also a change in the calculated errors in the physical solutions. We conclude that the value $\eta=1.00$ is the optimal choice.   

\section{Comparisons with the Cracow phase-shift analyses}

Both $\pi\pi$ phase-shift analyses KLR 97~\cite{kaminski97} and KLR 02~\cite{kaminski02} are based on their amplitude analysis of the CERN measurements on polarized target at low $t$ in 20 MeV mass bins presented in Ref.~\cite{kaminski97}. In Section II the Figure 1 shows that their two Solutions 1 (Up) and 2 (Down) for the moduli of the transversity amplitudes $|S_u|^2$ and $|S_d|^2$ are very close to our Solutions 1 and Solution 2, respectively, presented in the Figure 2. However as discussed in detail in  Ref.~\cite{svec12d} we differ in the relative phases $\Phi(L_u S^*_u)$ and $\Phi(L_d S^*_d)$. While in our phase-shift analysis these phases do not play any direct role they are important in their phase-shift analysis. Recall that $S_u \equiv g$ and $S_d \equiv h$.

In terms of our notation they assume
\begin{eqnarray}
f_S & = & N_S f \frac{\sqrt{q}}{m} \bigl( a_1 S_u +a_2 S_d)\\
\nonumber
f_P & = & N_P |A_{BW}(\rho^0)|e^{i\phi(\rho^0)}
\end{eqnarray}
where $N_S$ and $N_P$ are normalization factors, $a_1,a_2$ are complex kinematical parameters, $f$ is a vertex correction factor, $A_{BW}(\rho^0)$ is a Breit-Wigner amplitude at $\rho^0(770)$ and $\phi(\rho^0)$ is the its phase.
The parameters $a_1$ and $a_2$ satisfy constraint $|a_1|^2+|a_2|^2=1$ so that
\begin{eqnarray}
a_1 & = & \cos \alpha e^{i\theta_1}\\
\nonumber
a_2 & = & \sin \alpha e^{i\theta_2}
\end{eqnarray}

\begin{figure} [htp]
\includegraphics[width=12cm,height=10.5cm]{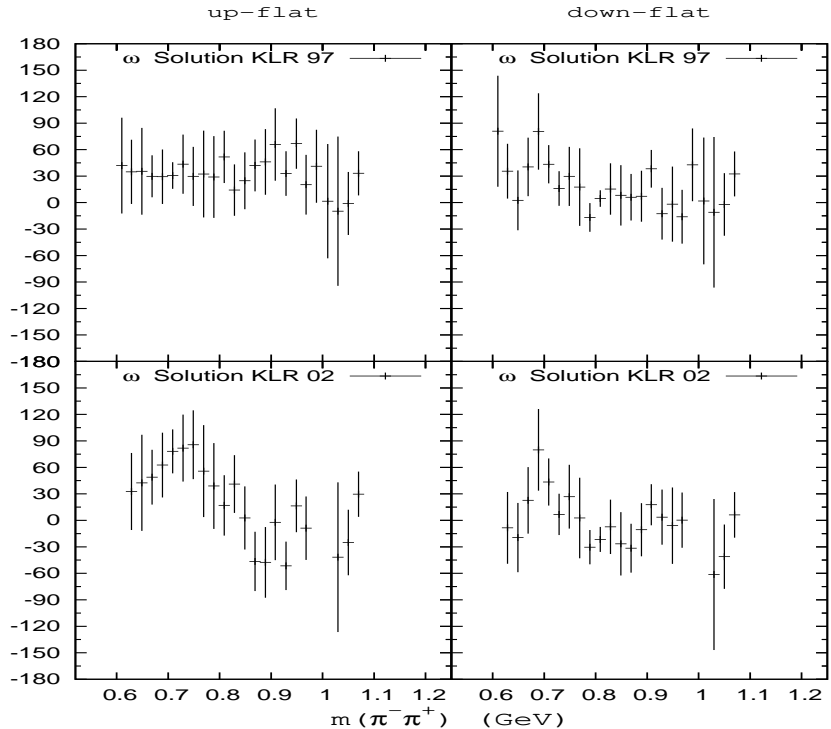}
\caption{Four Solutions for the relative phase $\omega=\Phi(S_d)-\Phi(S_u)$. Data from KLR 97~\cite{kaminski97} and KLR 02~\cite{kaminski02}.}
\label{Figure 20}
\end{figure}

This compares with our $N'_S=\frac{1}{G}$, $f'=\frac{1}{C_S}$ and
\begin{eqnarray}
a'_1 & = & \frac{1}{\sqrt{2}} e^{-i\theta_S-i\frac{\pi}{2}}\\
\nonumber
a'_2 & = &  \frac{1}{\sqrt{2}} e^{-i\theta_S+i\frac{\pi}{2}}
\end{eqnarray}
For $\sin \Theta_s \sim 0$ where $\Theta_s$ is the neutron scattering angle with respect to proton, we recover from relations (5.2) our relations (5.3). 

Their important assumption is that the absolute phases of the transversity amplitudes are given by an Ansatz
\begin{eqnarray}
\Phi(S_u) & = & \Phi(S_uL_u^*)+\phi(\rho^0)\\
\nonumber
\Phi(S_d) & = & \Phi(S_dL_d^*)+\phi(\rho^0)+\Delta
\end{eqnarray}
where $\Delta$ is a correction factor. Then $|f_S|^2$ will depend on the relative phase 
\begin{equation}
\omega = \Phi(S_d)-\Phi(S_u)=\Phi(S_dL_d^*)-\Phi(S_uL_u^*)+\Delta
\end{equation}
In the analysis KLR 97~\cite{kaminski97} it was assumed that $\Delta$ is a constant equal to $50.73^\circ$ below $K\bar{K}$ threshold. In the analysis KLR 02~\cite{kaminski02} $\Delta$ was determined at each mass bin from the BNL data on $\pi^- p \to \pi^0 \pi^0 n$ at 18.3 GeV/c~\cite{gunter01,pi0pi0pwa} using the more detailed relation (4.18) between the intensities $I^0$, $I$ and the transversity amplitudes. The results for this variable $\Delta$ are presented in Figure 4 of KLR 02~\cite{kaminski02} for both Solutions Up and Down of the transversity amplitudes. They were used to determine the new phase-shifts KLR 02 using the same transversity amplitudes as in the analysis KLR 97.

We used their data on $\Delta$ to reconstruct their relative phase $\omega$. Figure 20 shows the reconstructed $\omega$ for the constant $\Delta$ from KLR 97 and for the variable $\Delta$ from KLR 02. We note that the factors $a_g$ and $a_h$ in (4.18) depend explicitely on the vertex factor $C_\pi$ so that $\Delta$ and thus $\omega$ depend on the assumed $C_\pi$. 

In Ref.~\cite{svec12b} we have shown that $\omega$ can be determined at low $t$ analytically from the self-consistency condition of the bilinear terms of $S$ and $P$ wave transversity amplitudes. There are three solutions for $\omega$. The only solution consistent with the $\rho^0(770)$ resonant shape of $|L_1|^2$ and with the pion exchange dominance of $|S_1|^2$ requires $\cos \omega=-1$, or $\omega = \pm \pi$. All results for $\omega$ shown in Figure 20 are at variance with this exact result for $\omega$.

Apart from the zero structure of the phases $\Phi(S_dL_d^*)$ and $\Phi(S_uL_u^*)$ near 800 MeV the two amplitude analyses~\cite{svec12b,kaminski97} of the CERN data on polarized target are very similar. This is not surprising since both analyses use the same data set~\cite{rybicki96}. With $-t_{av}=0.066-0.068$(GeV/c)$^2$ in each 20 MeV mass bin of this data set the approximations (4.19) are well satisfied and we can use the relations (4.16) and (4.24). The main difference in the two $\pi\pi$ phase-shift analyses then are the equations (5.1) and (5.4) which define the amplitude $f_S$ differently from our definition. Apart from $C_S$ there are no other adjustable parameters in our definition of $f_S$ while the Cracow definition (5.1) involves in addition to $f$ the adjustable parameters $a_1, a_2$ and $\Delta$. The principal difference is in the relative phases $\omega$. Since the phase-shifts $\delta^0_S$ appear only mildly sensitive to $\omega$ the Solutions (2,2)1/ Down-flat KLR $97$ and the Solutions (2,2) joint/Down-flat KLR $02$ are mutually consistent.

\section{Conclusions and outlook}

In the Section II we have presented the $S$-wave transversity amplitudes from the 1997 Cracow amplitude analysis~\cite{kaminski97}, our amplitude Analysis I~\cite{svec12b} and the SMM Analysis~\cite{svec14a}. All these analyses present a convincing evidence for a rho-like structure at 770 MeV in transversity amplitudes in agreement with analyses of other polarized target measurements surveyed in Ref.~\cite{svec12d}. Since this structure is absent in $\pi^- p \to \pi^0 \pi^0 n$ it is interpreted as $\rho^0(770)-f_0(980)$ spin mixing. This evidence propagates into the $S$-wave spin mixing helicity amplitudes but is absent in the $S$-matrix helicity amplitudes. These helicity amplitudes are then related to the $\pi\pi$ scattering amplitudes.

All our phase-shift Solutions are analytical solutions in terms of the measured amplitudes and other data. The elastic analysis has two solutions for $\delta^0_S$. The solution in the joint analysis is unique but depends on the choice of the vertex factor $|C_\pi|^2$ for which we used the root $|C_\pi(-)|^2$ of the quadratic equation (4.37) assuming the Ansatz parameter $\eta=1.00$. This approximate solution of the system of equations (4.40)-(4.42) is acceptable since the calculated inelasticities have physical values below 1 and are not too far from the input value of $\eta$.

In the elastic analysis below the $K\bar{K}$ threshold our Solutions (2,2)1 and SpinMixing 1 are consistent with the 1997 Cracow Solution Down-flat KLR $97$. In the joint analysis of $\pi^-\pi^+$ and $\pi^0\pi^0$ data our Solutions (2,2) joint and SpinMixing joint are in a remarkable agreement with the 2002 Cracow Solution Down-flat KLR $02$. There is a similar remarkable agreement of our Solution (1,1) joint and the Cracow Solution Up-flat KLR $02$. In all these Solutions $\delta^0_S$ passes through $90^\circ$ at or near 770 MeV hinting at the signature of $\rho^0(770)-f_0(980)$ spin mixing in the $\pi\pi$ scattering amplitude $f_S$. 

There is no evidence of such mixing in both elastic and in both joint Solutions S-Matrix as expected from the $S$-matrix $\pi\pi$ scattering amplitudes. Crucially, all these Solutions are very similar suggesting the existence of a unique solution for $\delta^0_S$ and lending credence to its interpretation as genuine $S$-matrix amplitude $f_S$.

Our key observation is that our elastic and joint Solutions as well as the Cracow Solutions for the phase-shift $\delta^0_S$ presented in Figures 9-10 and in Figures 14-17 are consistent with the evidence for $\rho^0(770)-f_0(980)$ spin mixing in the $S$-wave transversity amplitudes from which all these Solutions ultimately arise. Our results suggest that a unique solution for the phase-shift $\delta^0_S$ for the $S$-matrix $\pi\pi$ scattering amplitudes with acceptable errors is attainable in future very high statistics measurements of $\pi^-p \to \pi^- \pi^+ n$ on polarized targets.

\end{document}